\documentstyle[12pt]{article}
\makeatletter

\@addtoreset{equation}{section}
\makeatother

\begin{document}
%%%%%%%%%%%%%%%%%%%%%%%%%%%%%%%%%%%%%%%%%%%%%%%%%%%%%%%%%%%%%%%%%%
%%%%%%%%%%%%%%%%%%%%%%%% Title %%%%%%%%%%%%%%%%%%%%%%%%%%%%%%%%%%%
%%%%%%%%%%%%%%%%%%%%%%%%%%%%%%%%%%%%%%%%%%%%%%%%%%%%%%%%%%%%%%%%%%
\begin{flushright}
DPUR-TH-24\\
DFPD11/TH/4\\
March, 2011\\
\end{flushright}
\vspace{20pt}
%\magnification=\magstep1
\pagestyle{empty}
\baselineskip15pt
%\font\cmssB=cmss17
%\font\cmssS=cmss10

\begin{center}
{\large\bf Free Differential Algebras and Pure Spinor Action in IIB 
Superstring Sigma Models
\vskip 1mm }

\vspace{20mm}
Ichiro Oda
          \footnote{
          E-mail address:\ ioda@phys.u-ryukyu.ac.jp} \\

\vspace{5mm}
           Department of Physics, Faculty of Science, University of the 
           Ryukyus,\\
           Nishihara, Okinawa 903-0213, Japan.\\

\vspace{5mm}

and

\vspace{5mm}

Mario Tonin
          \footnote{
          E-mail address:\ mario.tonin@pd.infn.it}\\
\vspace{5mm}
          Dipartimento di Fisica, Universita degli Studi di Padova,\\
          Instituto Nazionale di Fisica Nucleare, Sezione di Padova,\\
          Via F. Marzolo 8, 35131 Padova, Italy\\

\end{center}

%\maketitle

\vspace{5mm}
\begin{abstract}

In this paper we extend to the case of IIB superstring sigma models the 
method proposed in hep-th/10023500 to derive the pure spinor approach for type 
IIA sigma models. In particular, starting from the (Free) Differential 
Algebra and superspace parametrization  of type IIB supergravity, extended to 
include the BRST differential and all the ghosts, we derive the BRST 
transformations of fields and ghosts as well as the standard pure spinor 
constraints for the ghosts $\lambda $ related to supersymmetry. 
Moreover, using the method first proposed by us, we derive the 
pure spinor action for type IIB superstrings in curved supergravity 
backgrounds (on shell), in full agreement with the action first obtained by 
Berkovits and Howe.   

\end{abstract}

\newpage
\pagestyle{plain}
\pagenumbering{arabic}

\section{\large\bf Introduction}

Since bimillennium, Berkovits, sometimes together with 
collaborators,  developed a new formalism for superstrings 
 \cite{Ber1} - \cite{Ber8}, based on the concept of
pure spinor \cite{Nil}, \cite{Howe1}. It is a superspace approach, like the
Green-Schwarz (G-S) one, which however replaces the $\kappa$-symmetry of the 
G-S formulation with a BRST symmetry where the ghosts are pure spinors.  With 
respect to the G-S approach, it has the advantage to allow for a consistent,
super Poincar\`e invariant quantization of the superstrings in D=10 flat 
background or in special backgrounds as, for instance, $AdS_5\otimes S^5$
\cite{Ber7}. Moreover this formalism has the advantage over the RNS one to be
 able to treat fermions and R-R background fields in a natural way.
 
It is also interesting to extend the pure spinor approach to describe 
superstrings in curved supergravity backgrounds and write $\sigma$-models 
actions that are relevant, especially to deal with backgrounds in presence of
R-R fluxes. Here the seminal paper is \cite{BH}. In this paper, the authors
start from the more general classical action invariant under worldsheet 
conformal transformations and derive the supergravity constraints by requiring
nilpotence  of the BRST charge and holomorphicity of the BRST currents.
An equivalent approach is to require invariance of the action under BRST 
charge \cite{Cha1}.   

One could also reverse this procedure: i.e., start from the geometric 
formulation of the relevant, ten dimensional supergravity and then derive the
pure spinor action, as a modification of the G-S one, by requiring that the 
pure spinor action is BRST invariant. A first attempt to do that, 
restricted to the heterotic case, was done in \cite{Oda1}.   
Of course in order to be successful one should be  able to get from the 
geometric formulation of  the relevant supergravity model, the BRST
transformations of fields and ghosts. This can be done by generalizing a 
procedure well known for Yang-Mills theories \cite{Stora}, \cite{Bon1} and 
widely applied in gauge \cite{Zum}, \cite{Bon2}, \cite{MSZ}, 
topological \cite{Baul}, \cite{Fre1} and  supergravity theories \cite{Fre2}.
 We shall refer
to this procedure as the method of Extended (Free) Differential Algebra.
In addition, this method allows us to derive the pure spinor constraints for the
 ghosts under suitable conditions that are strongly related to 
the superembedding approach \cite{Super}. See also 
\cite{Oda2} where the relation between the pure spinor approach and the 
superembedding one was pointed out.

In \cite{tor1} a version of the method of Extended Differential Algebra was 
applied to the case of IIA superstring $\sigma$-models in order to derive the
ghost constraints and the BRST invariant action but the constraints obtained in
\cite{tor1} do not seem to fit with the standard pure spinor ones. In 
\cite{Ton}, one of the present authors (M.T.) has presented a variant of the 
method proposed in \cite{tor1} that allows to derive the standard pure spinor 
constraints and the pure spinor  action  for type IIA, D=10 superstring 
$\sigma$-models in full agreement with \cite{BH}.

In this paper we apply the approach of \cite{Ton} to get  the pure spinor 
constraints and the pure spinor action for the case of type IIB superstring
$\sigma$-models in 10 dimensions.

The paper is organized as follows. In Section 2, we review the geometrical 
superspace formulation of type IIB supergravity \cite{Sch}, \cite{HW},
\cite{CP}, \cite{Agat} and present the parametrization of torsions and 
curvatures by following \cite{HW}. In section 3, we write the G-S action, which 
is not a   trivial step as it could seem, since IIB supergravity (at the
classical level) is invariant under an $SL(2R)$ group, under which the 
NS-NS and R-R two superforms transform as a doublet. Therefore to write the 
G-S action we will adopt a method proposed in \cite{Berg} in a different 
context,  which preserves formally this $SL(2R)$ symmetry. Then the G-S action
suggests itself a rescaling (and a field redefinition) of the fields and 
superforms that describe type IIB supergravity. In section 4, we explain the 
method of Extended Differential Algebra and derive the pure spinor 
constraints for the ghosts.
Moreover we present the extended parametrization of rescaled torsions and 
curvatures in terms of the real and imaginary parts of the spinor-like 
vielbeins  and the other
complex fields, which  reduces the superspace parametrization of torsions and 
curvatures to a form very similar to that of IIA supergravity.  
Finally in section 5, we define the BRST charge and we give the BRST 
transformations of  the antighosts and the covariant momenta $d_{\underline{
\alpha}}$ (subsection 5.1) and we apply the method of \cite{Ton},
(and \cite{Oda1}), to derive the pure spinor action  of type IIB 
superstring $\sigma$-models in full agreement with \cite{BH} (section 5.2).
    
\section{\large \bf IIB Supergravity in 10 Dimensions}

\subsection{\bf Field Content and Notations}

 The D=10, IIB supergravity contains the following fields and forms:
the vector-like supervielbeins $E^{a} = dZ^{M}E_{M}{}^{a}(Z)$,
the spinor-like supervielbeins $(E^{\alpha}, E^{\ast \alpha})$,  
the two-superforms $ (B_{2}, B^{\ast}_{2})$,  
the four-superform $C_{4}$, 
the chiral spinors $(\Lambda_{\alpha}, \Lambda^{\ast}_{\alpha})$,
the Lorentz superconnection $ \Omega^{ab}$,  
and the scalars $( V_{+}{}^{i},V_{-}{}^{i})$.    

$V_{\pm}{}^{i}$ belong to the coset $SL(2R)/U(1)$ transforming as 
$ V'_{\pm}{}^{i} =
e^{\pm 2 i \epsilon}V_{\pm}{}^{j}\Lambda_{j}{}^{i}$ with $\Lambda_{j}{}^{i} \in
SL(2R)$
and $V_{\pm}^{i} $ satisfy the reality condition
\begin{eqnarray}
V^{i}_{+} = \bar V^{i}_{-},  \qquad V^{i}_{-} = \bar V^{i}_{+}, 
\label{14}
\end{eqnarray}
where, if $\psi^{i}$ is an $SL(2R)$ doublet, we define
\begin{eqnarray}
\bar \psi^{i} = (\tau_{1}\psi^{\ast})^{i}.
\nonumber
\end{eqnarray}
Moreover $$ \epsilon_{ij}V_{+}{}^{i}V_{-}{}^{j} = 1, \qquad
(V_{-}{}^{i}V_{+}{}^{j}) - (V_{-}{}^{j}V_{+}{}^{i}) = - 2 \epsilon^{ij}.$$
Then one defines the one-superforms $$ 2i Q = \epsilon_{ij}V_{-}{}^{j} d V_{+}
{}^{i},$$ which plays the role of $U(1)$ connection and $$ R_{1} = 
  \epsilon_{ij}V_{+}{}^{i} d V_{+}{}^{j}, \qquad R^{\ast}{}_{1} = 
\epsilon_{ij}V_{-}{}^{j} d V_{-}{}^{i}. $$
% In our conventions, derivatives act from right to left. 
%Moreover objects with even grading  always commute and objects with odd 
%grading anticommute among themselves and commute with those of even grading.

With respect to the structure group $U(1)$, $V_{\pm}{}^{i}$ have charges 
$ q = \pm 2 $,   $(E^{\alpha}, E^{\ast \alpha})$ have charges $q = \pm 1$, 
the two-superforms $ (B_{2}, B^{\ast}_{2})$ have charges $q = \pm 2$, 
 $(\Lambda_{\alpha}, \Lambda^{\ast}_{\alpha})$ have charges
$ q = \pm 3$ and $ (R_{1}, R^{\ast}_{1}) $ have charges $q= \pm 4$.
$ E^{a}$, $\Omega^{ab}$ and $C_{4}$ are uncharged.
Covariant derivatives involve the Lorentz connection $\Omega^{ab} = -
 \Omega^{ba}$ acting on Lorentz 
tensors and spinors and the connection $Q$ acting on charged fields. It is 
also convenient to  introduce, at no cost, a Weyl connection $\Omega$ with 
zero curvature such that spinors with $p$ upper and $q$ lower 
spinorial indices have Weyl charge $p - q$. Since the Weyl curvature vanishes,
$\Omega$ is a pure gauge.  For instance,
$$ T^{a} = \Delta E^{a} = dE^{a} + E^{b}\Omega_{b}{}^{a},$$
$$ T^{\alpha} = \Delta E^{\alpha} = d E^{\alpha} + E^{\beta} {1 \over 4}
(\Gamma^{ab})_{\beta}{}^{\alpha}\Omega_{ab} + E^{\alpha} \Omega -
i  E^{\alpha}Q, $$ $$ \Delta \Lambda_{\alpha} = 
d \Lambda_{\alpha} + {1 \over 4}(\Gamma^{ab})_{\alpha}{}^{\beta}\Omega_{ab}
\Lambda_{\beta} + \Omega \Lambda_{\alpha}  - 3 i  \Lambda_{\alpha}Q,$$ etc. 
The Lorentz curvature is as usual  
$$ R^{ab} = d \Omega^{ab} + \Omega^{a}{}_{c}\Omega^{cb}.$$
Instead of the charged two-superforms $(B_{2 },B^{\ast}_{2 })$ it is 
convenient 
to use the uncharged ones $B_{2}^{i}$ which transform as a doublet 
of $SL(2R) $ and are defined as 
\begin{eqnarray}
 B_{2}^{i} = V_{-}^{i}B_{2} + V_{+}^{i}B^{\ast}_{2}.
\label{13}
\end{eqnarray}
 The curvature of $B_{2}^{i}$ is 
$$ H_{3}^{i}= d B_{2}^{i},$$  and 
the curvature of $C_{4}$ is $$ F_{5} = i d C_{4} + 2 i \epsilon_{ij}B_{2}^{i}
d B_{2}^{j}. $$ 
Notice that in our notations  $F_{5}$ is purely imaginary.

These torsions and curvatures satisfy the Bianchi identities
\begin{eqnarray}
\Delta T^{a} = E^{b}R_{b}{}^{a},
\label{11}
\end{eqnarray}
\begin{eqnarray}
\Delta T^{\alpha} = E^{\beta}R_{\beta}{}^{\alpha} + {1 \over 2} R_{1}
R^{\ast}_{1} E^{\alpha},
\label{12a}
\end{eqnarray}

\begin{eqnarray}
\Delta T^{\ast\alpha} = E^{\ast\beta}R_{\beta}{}^{\alpha} -{ 1 \over 2} R_{1}
R^{\ast}_{1} E^{\ast \alpha},
\label{12b}
\end{eqnarray}
where $ R_{\alpha}{}^{\beta} ={1 \over 4} R^{ab}(\Gamma_{ab})_{
\alpha}{}^{\beta}.$
 Moreover
%%%%%%%%
\begin{eqnarray}
\Delta R^{ab} = 0, 
\end{eqnarray}
%%%%%%%%%%%%%
\begin{eqnarray}
 dH_{3}^{i} = 0,  
\end{eqnarray}
%%%%%%%%%%%%%%%%%%%%%%%
\begin{eqnarray}
dF_{5} = - 2 i \epsilon_{ij} H_{3}^{i}H_{3}^{j}, 
\end{eqnarray}
%%%%%%%%%%%%%%%%%%%
\begin{eqnarray}
dQ = { i \over 2}R_{1}R^{\ast}_{1}. 
\label{10}
\end{eqnarray}

\subsection {\bf Parametrization of Torsions and Curvatures}

We shall adopt the parametrization of Howe and West (H-W) \cite{HW},
modulo some different conventions. For H-W, complex conjugation reverses the order of
 the factors, while for us it is simply to take the complex conjugate without reversing the 
order. Moreover in H-W if X is odd $ E^\alpha X = - X  E^\alpha $ and 
$  ({E^{\alpha}  E^{\beta}})^{\ast} = -  E^{\ast \alpha} E^{\ast \beta}. $   
 In our conventions, derivatives act from right to left. 
Moreover objects with even grading  always commute and objects with odd grading
 anticommute among themselves and commute with those of even grading.

If we denote with $X_{0}$ the objects in the notations of H-W, the 
dictionary among our notation and that of H-W is the following:
$$ E^{\alpha} = e^{{-i\pi \over 4}} E_{0}^{\alpha}, \qquad 
 E^{\ast \alpha} = e^{{-i\pi \over 4}} E_{0}^{\ast \alpha},$$ 
$$ \Lambda^{\alpha} = e^{{i\pi \over 4}} \Lambda_{0}^{\alpha}, \qquad 
 \Lambda^{\ast \alpha} = e^{{i\pi \over 4}} \Lambda_{0}^{\ast \alpha},$$ 
$$ B_{2} =  B_{0 2},  \qquad  C_{4} = - i C_{0 4 }.$$ 
Then one gets the following parametrization 
%%%%%%%%%%%%%%%%%%%%%%%%
\begin{eqnarray}
T^{a} = (E^{\ast}\Gamma^{a}E),
\label{1}
\end{eqnarray}
%%%%%%%%%%%%%%%%%%%%%%%
\begin{eqnarray}
T^{\alpha} = [{1 \over 2} (E^{\ast}\Gamma^{a}E^{\ast})(\Gamma_{a}\Lambda)^{
\alpha} - E^{\ast \alpha}(E^{\ast} \Lambda)] + {1 \over {16}} E^{a}[ 
3 (E^{\ast}\Gamma^{bc})^{\alpha}H_{abc} +
{1 \over 3} (E^{\ast}\Gamma_{abcd})^{\alpha}H^{bcd}]
\nonumber
\end{eqnarray}
\begin{eqnarray}
 + E^{c}E^{\beta}[ {9}
\chi_{c}\delta_{\beta}{}^{\alpha} + {3 \over 2} (\Gamma^{c}\Gamma^{b})_{\beta}
{}^{\alpha}
\chi^{b} + { 1 \over 2}(\Gamma^{ab})_{\beta}{}^{\alpha}\chi_{abc} + 
{1 \over 4}(\Gamma_{c}\Gamma_{abd})_{\beta}{}^{\alpha}\chi^{abd}
\nonumber
\end{eqnarray}
\begin{eqnarray}
 + {1 \over {48}} (\Gamma^{abde})_{
\beta}{}^{\alpha}({1 \over 4} F^{(+)}_{abdec} +  \chi^{(-)}_{abdec}) ] +
 {1 \over 2}E^{a}E^{b}T_{ba}^{\alpha}, 
\label{2}
\end{eqnarray}
%%%%%%%%%%%%%%%%%%%%%%%%%
%\underline{N.B.} $$   {1 \over 2} (E^{\ast}\Gamma^{a}E^{\ast})(\Gamma_{a}
%\Lambda)^{\alpha} - E^{\ast \alpha}(E^{\ast} \Lambda) = - {1 \over 4} [
%{1 \over 2} (\Gamma^{ab} E^{\ast})_{ \alpha}(E^{\ast}\Gamma_{ab}\Lambda) -
%E^{\ast \alpha} (E^{\ast}\Lambda)] $$  
\begin{eqnarray}
T^{\ast \alpha} =  [{1 \over 2}(E\Gamma^{a}E)(\Gamma_{a}
\Lambda^{\ast})^{\alpha} -  E^{\alpha}(E \Lambda^{\ast})] + {1 \over {16}} 
E^{a}[3 (E \Gamma^{bc})^{\alpha}
H^{\ast}_{abc} + {1 \over 3} (E\Gamma^{abcd})^{\alpha}H^{\ast}_{bcd}]
\nonumber
\end{eqnarray}
\begin{eqnarray}
 +  E^{c}E^{\ast \beta}[ - {9}\chi_{c}\delta_{\beta}{}^{\alpha} - {3 \over 2} 
(\Gamma_{c}\Gamma_{b})_{\beta}
{}^{\alpha}\chi^{b} + {1 \over 2} (\Gamma^{ab})_{\beta}{}^{\alpha}\chi_{abc} + 
{1 \over 4}(\Gamma_{c}\Gamma_{abd})_{\beta}{}^{\alpha}\chi^{abd}
\nonumber
\end{eqnarray}
\begin{eqnarray}
 - {1 \over {48}} (\Gamma^{abde})_{\beta}{}^{\alpha}({1 \over 4 }F^{(+)}_{
abdec} +  \chi^{(-)}_{abdec}) ] + {1 \over 2}
E^{a}E^{b}T^{\ast \alpha}_{ba},  
\label{3}
\end{eqnarray}
%%%%%%%%%%%%%%%%%%%%%%%%% 
%\underline{N.B.} $$   {1 \over 2} (E \Gamma^{a} E )(\Gamma_{a}
%\Lambda^{\ast})^{ \alpha} - E^{\alpha}(E  \Lambda^{\ast}) = - {1 \over 4} [
%{1 \over 2} (\Gamma^{ab} E)^{\alpha}(E \Gamma_{ab}\Lambda^{\ast}) -
%E^{\alpha} ( E \Lambda^{\ast})] $$ 
\begin{eqnarray}
R^{ab} =  3 (E^{\ast}\Gamma^{abc}E)\chi_{c} - 2 (E^{\ast}\Gamma_{c}E)
\chi^{abc} + {1 \over 2}
(E^{\ast} \Gamma^{a}\Gamma^{cde}\Gamma^{b}E) \chi_{cde} + {1 \over 6}(
E^{\ast}\Gamma_{cde}E)( {1 \over 4} F^{(+)abcde} + \chi^{(-) abcde})
\nonumber
\end{eqnarray}
\begin{eqnarray}
 - {1 \over 4}  [ (E\Gamma_{c}E)H^{\ast abc} +
 (E^{\ast}\Gamma_{c}E^{\ast})H^{abc} ] + {1 \over{48}} [(E\Gamma^{a}\Gamma^{
cde}\Gamma^{b}E) H^{\ast}_{cde} +
(E^{\ast}\Gamma^{a}\Gamma^{cde}\Gamma^{b}E^{\ast}) H_{cde}]
\nonumber
\end{eqnarray}
\begin{eqnarray}
 + {1 \over 2}  E^{c}[ (E^{\ast}\Theta_{c}^{ab}) +  
(E \Theta^{\ast ab }_{c})] + {1 \over 2}  E^{c}E^{d}R_{dc}{}^{ab}, 
\label{4}
\end{eqnarray}
%%%%%%%%%%%%%%%%%%%%%%%%%%%%
where $$ \Theta^{ab}_{\alpha c}= (\Gamma_{c}T^{ab})_{\alpha} - 2 (\Gamma^{[a}
T^{b]}{}_{c})_{\alpha}, $$
$$ \chi^{(r)} = {1 \over {16}} (\Lambda \Gamma^{(r)}\Lambda^{\ast}). $$
Notice that in our notation $ \chi_{a}$ and $\chi_{abcde}$ 
are purely imaginary whereas $\chi_{abc}$ is real.

Moreover
%%%%%%%%%%%%%%%%%%%%%%
\begin{eqnarray}
\Delta V_{+}^{i} = V_{-}^{i}[ -2(E\Lambda) + E^{a}R_{a}],
\nonumber
\end{eqnarray}
\begin{eqnarray}
\Delta V_{-}^{i} = V_{+}^{i}[ - 2(E^{\ast}\Lambda^{\ast}) + E^{a}R^{\ast}_{a}],
\label{5}
\end{eqnarray}
%%%%%%%%%%%%%%%
\begin{eqnarray}
H_{3}^{i} = + {1 \over 2}[ E^{c}(E\Gamma_{c}E)V_{-}^{i} +
 E^{c}(E^{\ast}\Gamma_{c}E^{\ast})V_{+}^{i}] +  {1 \over 2}E^{a}E^{b}[ 
(E^{\ast}\Gamma_{ba}\Lambda)V_{-}^{i} + (E\Gamma_{ba}\Lambda^{\ast})
V_{+}^{i}]
\nonumber
\end{eqnarray}
\begin{eqnarray}
 + {1 \over 6 } E^{a}E^{b}E^{c}[H_{abc}V_{-}^{i} + H^{\ast}_{abc}V_{+}^{i}],
\label{7}
\end{eqnarray} 
%%%%%%%%%%%%%%%%%%%%%%%%%
\begin{eqnarray}
F_{5} = {1 \over 6} E^{a}E^{b}E^{c}(E^{\ast}\Gamma_{cba}E)  + {1 \over {5!}}
E^{a}E^{b}E^{c}E^{d}E^{e}F_{abcde},
\label{8}
\end{eqnarray}
%%%%%%%%%%%%%%%%%%%%%%%%%
and
%%%%%%%%%%%%%%%%%%%%

\begin{eqnarray}
\Delta \Lambda_{\alpha} =  {1\over 2} R^{a}(\Gamma_{a}E^{\ast})_{\alpha} +
{1 \over {24}} (\Gamma^{abc}E)_{\alpha}H_{abc} + E^{b}\Delta_{b}
\Lambda_{\alpha},
\nonumber
\end{eqnarray}
\begin{eqnarray}
\Delta \Lambda^{\ast}_{\alpha} =   {1 \over 2}R^{\ast a}(\Gamma_{a}E)_{\alpha}
 + {1 \over {24}} (\Gamma^{abc}E^{\ast})_{\alpha}
H^{\ast}_{abc} + E^{b}\Delta_{b}\Lambda^{\ast}_{\alpha}.
\label {6}
\end{eqnarray}
%%%%%%%%%%%%%%%%%%%%%%%%%
If $Z_{a_{1}...a_{5}}$ is a 5-indexed superfield, $$ Z^{(\pm)}{}_{a_{1}...
a_{5}} = {1 \over 2} ( Z_{a_{1}...a_{5}} \pm (*Z)_{a_{1}...a_{5}}), $$ are
 its self-dual and antiself-dual components.
$ \chi_{a_{1}...a_{5}}$ is antiself-dual, i.e., $ \chi_{a_{1}...a_{5}} = 
 \chi^{(-)}_{a_{1}...a_{5}}$. Moreover $$  F^{(-)}_{a_{1}...a_{5}} = - 8 
 \chi_{a_{1}...a_{5}},$$
and
$$ Z_{abcde} = {1 \over 192} [ F_{abcde} + 12 \chi_{abcde}] = {1 \over 192}
[F^{(+)}_{abcde} + 4 \chi^{(-)}_{abcde}]. $$
{}From the definitions of $R_{1}$ , $R^{\ast}_{1}$ and Q one has
\begin{eqnarray}
R_{1} = -2 (E\Lambda) + E^{a}R_{a}, 
\nonumber
\end{eqnarray}
\begin{eqnarray}
R^{\ast}_{1} = - 2 (E^{\ast}\Lambda^{\ast}) + E^{a}R^{\ast}_{a}. 
\label{9}
\end{eqnarray}

%%%%%%%%%%%%%
 
\section{\bf Green-Schwarz Action, Rescaling  and Field Redefinition}

\subsection{\bf G-S Action}

As a first step to obtain the pure spinor action, one must write the G-S 
action for the IIB sigma model. Since $B_{2}^{i}$ is an 
$SL(2R)$ doublet
and since the W-Z term involving   $B_{2}^{i}$ must be real and must have 
a scalar structure, we will introduce a complex, constant $SL(2R)$ doublet 
$q_{i}$ (with $ \bar q_{i} = (\tau_{1}q^{\ast})_{i} $ ) \cite{Berg}. 

Notice that from the reality condition (\ref{14}) and (\ref{13}) one has
 $ H_{3}^{i} = \bar H_{3}^{i}$ so that also 
$ B_{2}^{i} = \bar B_{2}^{i}.$ 
Therefore,  writing $ n^{i} = {1 \over 2}( q_{i} + \bar q_{i})$, one has that
 $$ {1 \over 2}[q_{i}B_{2}^{i} + \bar  q_{i} \bar B_{2}^{i}] = n_{i}
B_{2}^{i},$$ is real and ``scalar''. Moreover $ n_{i} = \bar n_{i}.$
Then by defining
%%%%%%%%%%%%%%%%%%%%%%%%%%%%%
\begin{eqnarray}
e^{2\phi} = n_{i}V_{-}^{i},   \qquad 
e^{2\phi^{\ast}} = (n_{i}V_{-}^{i})^{\ast} = \bar n_{i}V_{+}^{i},
\label{15}
\end{eqnarray}
one has
$$    n_{i}B_{2}^{i} = e^{2\phi} B_{2} + e^{2\phi^{\ast}} B_{2}^{\ast}. $$

Now we propose the following G-S action (in the conformal gauge):
\begin{eqnarray}
I_{GS} = {1 \over 2} \int [  e^{\phi}e^{\phi^{\ast}}E_{+}^{a}E_{- a } + 
2 n_{i}B_{2}^{i}]
\nonumber
\end{eqnarray}
\begin{eqnarray}
 = \int [ {1 \over 2} e^{\phi}e^{\phi^{\ast}}E_{+}^{a}E_{- a } + 
e^{2\phi}B_{2} + e^{2 \phi^{\ast}}B_{2}^{\ast}].
\label{16}
\end{eqnarray}
The factor $ e^{\phi}e^{\phi^{\ast}} $ in front of $ E_{+}^{a}E_{- a }$ will 
become clear in the following.

\subsection{\bf Rescaling}

 The variation of the second term of $I_{GS}$ involves the 3-superform 
$H_{3} \equiv  n_{i}H_{3}^{i} $ which, according to (\ref{7}) and (\ref{15}), 
is 

\begin{eqnarray}
H_{3} = + {1 \over 2}[ E^{c}(E\Gamma_{c}E)e^{2\phi} +
 E^{c}(E^{\ast}\Gamma_{c}E^{\ast})e^{2 \phi^{\ast}}] +  {1 \over 2}E^{a}E^{b}[ 
(E^{\ast}\Gamma_{ba}\Lambda)e^{2\phi} + (E\Gamma_{ba}\Lambda^{\ast})
e^{2\phi^{\ast}} ]
\nonumber
\end{eqnarray}
\begin{eqnarray}
+ {1 \over 6 } E^{a}E^{b}E^{c}[H_{abc}e^{2 \phi} + H^{\ast}_{abc}e^{2 
\phi^{\ast}}].
\label{7bis}
\end{eqnarray}  

Eq. (\ref{7bis}) suggests  to  perform the following rescaling
\begin{eqnarray}
\epsilon^{\alpha} ={e^{\phi} \over e^{k(\phi + \phi^{\ast})}} E^{\alpha},  
\qquad \epsilon^{\ast \alpha} = {e^{\phi^{\ast}} \over e^{k(\phi + 
\phi^{\ast})}} E^{\ast \alpha},
\nonumber
\end{eqnarray} 

\begin{eqnarray}
{\cal E}^{a} =  e^{2 k(\phi + \phi^{\ast})} E^{a},
\nonumber
\end{eqnarray}

\begin{eqnarray}
 \pi_{\alpha} =  \frac{e^{3\phi}}{e^{(3 k + 1)( \phi + \phi^{\ast})} }
\Lambda_{\alpha}, \qquad \pi^\ast_\alpha =  \frac{e^{3 \phi^{\ast}}} 
{e^{(3k +1)( \phi + \phi^{\ast})}} \Lambda^\ast_{\alpha}.
\nonumber
\end{eqnarray}
The simplest choice $ k = 0 $ yields unpleasant factors   $ {1 \over {e^{\phi}
e^{\phi^{\ast}}}} $ in front of the torsions and curvatures expressed in 
terms of the rescaled fields. These unpleasant factors can be removed by 
choosing $ k = {1 \over 4}$. Then
\begin{eqnarray}
\epsilon^{\alpha} ={e^{\phi} \over e^{{1 \over 4}(\phi + \phi^{\ast})}}
E^{\alpha},  
\qquad \epsilon^{\ast \alpha} = {e^{\phi^{\ast}} \over e^{{1 \over 4}(\phi + 
\phi^{\ast})}} E^{\ast \alpha},
\label{22}
\end{eqnarray} 

\begin{eqnarray}
{\cal E}^{a} =  e^{{1 \over 2}(\phi + \phi^{\ast})} E^{a},
\label{24}
\end{eqnarray}

\begin{eqnarray}
 \pi_{\alpha} =  {{e^{3\phi}} \over {e^{{7 \over 4} ( \phi + \phi^{\ast})}}}
\Lambda_{
\alpha}, \qquad \pi^\ast_{\alpha} =  {{e^{3 \phi^{\ast}}} \over {
e^{{7 \over 4}( \phi + \phi^{\ast})}}} \Lambda^\ast_{\alpha},
\label{25}
\end{eqnarray} 
and therefore

\begin{eqnarray} 
\kappa^{(r)} = {1 \over {e^{{1 \over 2}( \phi + \phi^{\ast})}}} \chi^{(r)} 
\equiv {1 \over {16}} (\pi \Gamma^{(r)} \pi^\ast),
\label{25a}
\end{eqnarray}

\begin{eqnarray}
{\cal H}_{abc} = { e^{2\phi} \over e^{{3 \over 2}(\phi + \phi^{\ast})}}
 H_{abc},
\qquad  {\cal H}_{abc}^{\ast} = { e^{2\phi^{\ast}}\over e^{{3 \over 2}(\phi + 
\phi^{\ast})}} H_{abc}^{\ast}.
\label{27}
\end{eqnarray}

Moreover, from (\ref{5}) and (\ref{8}), 
\begin{eqnarray}
 \rho^{a} = { {e^{4 \phi}} \over {e^{{5 \over 2}(\phi  + \phi^{\ast})}} }  R^{a},
\qquad \bar \rho^{a} = { {e^{4 \phi^{\ast}}} \over { e^{{5 \over 2}(\phi + 
\phi^{\ast})}}}  \bar R^{a},
\label{28}
\end{eqnarray}
and
\begin{eqnarray}
{\cal F}^{(+)}_{abcde} =  e^{{5 \over 2}(\phi + \phi^{\ast})} F^{(+)}_{abcde}.
\label{28bis}
\end{eqnarray}
 Notice that $B_{2}^{i}$, $C_{4}$ and $\Omega^{ab}$ (and therefore
$H_{3}^{i}$, $F_{5}$  and $R_{a}{}^{b}$) are not rescaled.

\subsection {\bf Field redefinitions}

Moreover from (\ref{5}) one has 
\begin{eqnarray}
\Delta \phi =  - (\epsilon^{\ast}
\pi^{\ast}) + {1 \over 2}{\cal E}^{a}\rho^{\ast}_{a},
\label{29a}
\end{eqnarray}
\begin{eqnarray}
\Delta \phi^{\ast} =  - (\epsilon
\pi) + {1 \over 2} {\cal E}^{a}\rho_{a},
\label{29b}
\end{eqnarray}
and the torsion of the rescaled vielbein ${\cal E}^{a}$ becomes 
\begin{eqnarray}
\Delta {\cal E}^{a} = (\epsilon^{\ast} \Gamma^{a}\epsilon) + 
[{1 \over 2}(\epsilon \pi) + {1 \over 2}(\epsilon^{\ast} \pi^{\ast})]
{\cal E}^{a} + {1 \over 4} {\cal E}^{a}{\cal E}^{b}(\rho_{b} + 
\rho^{\ast}_{b}), 
\label{30a}
\end{eqnarray}
which does not vanish in the sectors (1,1) and (2,0) \footnote{Once a supervielbein basis is fixed, any $n$-superform $\psi_{n}$
can be decomposed as $ \psi_{n} = \sum \psi_{(p,q)} $, ($p + q = n$), where
$\psi_{(p,q)} $ is the component of $\psi_{n}$ proportional to $p$
vector-like and $q$ spinor-like vielbeins. Then $\psi_{(p,q)}$ is called the
$(p,q)$ sector of $\psi_{n}$.}.
It is convenient to perform a redefinition of the spinor-like vielbeins and 
of the Lorentz connection so that $ \Delta'{\cal E}^{a}$ vanishes in the 
sectors (1,1) and (2,0) (where now $\Delta'$ denotes the covariant 
differential given in terms of the redefined connection ${\Omega'}^{ab}$).
Indeed, calling ${\cal E}$ the redefined spinor-like vielbeins, let us 
consider the transformations
\begin{eqnarray}
{\cal E}^{\alpha} = \epsilon^{\alpha} - {1 \over 2} {\cal E}^{b}(\Gamma_{b}
\pi^{\ast})^{\alpha}, \qquad {\cal E}^{\ast \alpha} = \epsilon^{\ast \alpha} - 
{1 \over 2} {\cal E}^{b}(\Gamma_{b}\pi)^{\alpha},
\label{31}
\end{eqnarray}
\begin{eqnarray}
{\Omega'}^{ab} = \Omega^{ab} + \delta \Omega^{ab},
\label{32a}
\end{eqnarray}
where
\begin{eqnarray}
\delta \Omega^{ab} = {1 \over 2} [ ({\cal E}\Gamma^{ab}\pi) +   
 ({\cal E}^{\ast}\Gamma^{ab}\pi^{\ast})] + 4 {\cal E}^{c}\kappa_{c}{}^{ab} - 
{1 \over 2}{\cal E}^{c} \delta_{c}^{[a}(\rho^{\ast} + \rho)^{b]},
\label{32b}
\end{eqnarray}
and $\pi , \pi^{\ast}$ and $\rho, \rho^{\ast}$ are defined in (\ref{25}),
(\ref{28}) and $\kappa^{abc}$ is defined in (\ref{25a}) with $\Gamma^{(r)} = 
\Gamma^{abc}$.
Then one can immediately verify that under the transformations (\ref{31}),
(\ref{32a}) and (\ref{32b}) 
\begin{eqnarray}
\Delta' {\cal E}^{a} = ({\cal E}^{\ast}\Gamma^{a}{\cal E}).
\label{33}
\end{eqnarray}
It is then straightforward to compute the torsions and curvatures with the 
rescaled and redefined fields and forms. 
However before doing that, let us introduce the associated Extended Free 
Differential Algebra with its related torsions and curvatures.\\
 
\section{\bf Extended Free Differential Algebra} 

\subsection{\bf BRST differential, extended forms and ghosts}

A convenient way to get the BRST transformations of fields and ghosts is to 
consider the Extended Differential Algebra that amounts to the following 
recipe:

a) Define the hatted quantities as 
\begin{eqnarray}
 \hat d = d + s + \delta \equiv d + \tilde s + \tilde \delta, 
\label{330}
\end{eqnarray}
\begin{eqnarray}
\hat{\cal E}^{a} = {\cal E}^{a} + \lambda^{a},
\label{331}
\end{eqnarray}
$$\hat {\cal E}^{\alpha}  = {\cal E}^{\alpha} + 
\lambda^{\alpha}, $$
\begin{eqnarray}
 \hat {\cal E}^{\ast \alpha}=   
{\cal E}^{\ast \alpha} + \lambda^{\ast \alpha},  
\label{332}
\end{eqnarray}
\begin{eqnarray} 
\hat \Omega_{ab} =\hat {\cal E}^{C} {\Omega'}_{C}{}^{ab} + \psi^{ab} \equiv 
{\Omega'}^{ab} + \tilde \psi^{ab},
\label{333}
\end{eqnarray}
\begin{eqnarray} 
\hat B^{i}_{2} = \hat {\cal E}^{A}\hat {\cal E}^{B}B^{i}_{BA} + 
\sigma^{i}_{1} \equiv  B^{i} + \tilde \sigma^{i}_{1},
\label{334}
\end{eqnarray}
\begin{eqnarray} 
\hat C_{4} = \hat {\cal E}^{A_{1}} \cdots \hat {\cal E}^{A_{4}}
C_{A_{1} \cdots A_{4}} + \sigma_{3}
\equiv  C_{4}  + \tilde \sigma_{3},
\label{335}
\end{eqnarray}  
where the ghosts $\lambda^{a}$, $\lambda^{\alpha}$, $\lambda^{\ast \alpha}$, $\psi^{ab}$,
$\sigma^{i}_{1}$ and $\sigma_{3}$ have ghost number $n_{gh}= 1$ but 
$\sigma^{i}_{1}$ and $\sigma_{3}$, being one-form and 3-form respectively,
contain ghosts of ghosts of higher ghost number.
Since the Green-Schwarz action involves only the two-superform $n_{i}B^{i}_{2}$
for our purposes the relevant ghost related to $B_{2}^{i}$ is 
$(n_{i}\sigma^{i}_{1})$. 

b) Assume that the ghost $\lambda^{a}$ related to ${\cal E}^{a}$ and the ghost
$n_{i}\tilde \sigma^{i}_{1}$ 
vanish so that 
\begin{eqnarray}
\hat {\cal E}^{a} = {\cal E}^{a}, 
\label{336}
\end{eqnarray}
\begin{eqnarray}
  n_{i}\hat B^{i}_{2} = n_{i} B^{i}_{2}. 
\label{337}
\end{eqnarray}
c) Write  the extended parametrization for hatted torsions and 
curvatures simply copying that of the unhatted ones.

Following \cite{Bon1}, one can give a geometric interpretation to Eqs. 
(\ref{330}) - (\ref{335}) by adding an odd, non-dynamical dimension to the
superspace, with odd 
coordiante $\eta$ . Then the ghosts (and ghosts of ghosts) in Eqs. (\ref{331}) -
(\ref{335}) can be identified with the components of the corresponding hatted
superforms along this odd direction and $ s $ in (\ref{330}) as the 
differential along $\eta$.   

A justification of the assumption expressed under the point b) can be done in 
the framework of the superembedding approach. In this approach the w.s. is 
considered a super w.s. and the ghosts which arise in the definitions of the
hatted superforms are just the pull-back of these extended superforms along an
odd super w.s. direction, let say, of odd coordinate $\kappa $. Then the 
condition $\lambda_{ws}^{a}= d \kappa \partial_{\kappa} Z^{M}{\cal E}_{M}^{a}=
0 $ is just the 
fundamental constraint of the superembedding approach, i.e., the requirement that 
the pull-back of the vector-like vielbeins along an odd super w.s. direction,
 vanishes. On this line,  the  condition $ (n_{i} \tilde \sigma^{i}_{ws 1}) =
 d \kappa \partial_{\kappa} Z^{M}{\cal E}_{M}^{A}{\cal E}^{B}B_{BA} = 0, $  
together with $\lambda_{ws}^{a}= 0 $ expresses the fact that if, as in 
reonomic approach, one writes the G-S action as the integral of a top 2-form 
in the extended superspace, the pullback of this top 2-form 
vanishes along the odd super w.s. direction $\kappa$.

However, $\lambda^{a}$ and $\lambda_{ws}^{a}$ (as well as  
$ (n_{i} \tilde \sigma^{i}_{ 1})$ and $ (n_{i} \tilde \sigma^{i}_{ws 1})$)
are {\it a priori} different objects since  $\lambda^{a}$ and $ (n_{i} \tilde 
\sigma^{i}_{ 1})$ are the odd components of $\hat {\cal E}^{a}$
and $ n_{i}\hat B_{2}^{i}$ in the extended superspace ($Z^{M}, \eta $) whereas  
 $\lambda_{ws}^{a}$ and $ (n_{i} \tilde 
\sigma^{i}_{ws 1})$ are the odd components of the pull-back of
 $\hat {\cal E}^{a} $ and $ n_{i}\hat B_{2}^{i}$ in the extended w.s. 
($\xi^{i}, \kappa$). They can be identified if one specifies the odd part
of the embedding of the extended w.s. on the extended superspace. Indeed,
if $Z^{M}(\xi, \kappa) = Z^{M}(\xi) + \kappa Y^{M}(\xi)$ is this embedding,
one can choose $ \partial_{\kappa}Z^{M} \equiv Y^{M}= {\cal E}_{\eta}^{A}
{\cal E}_{A}^{M} $ so that $\lambda^{a} = \lambda_{ws}^{a}$
 and $n_{i} \tilde \sigma^{i}_{ 1} = n_{i} \tilde \sigma^{i}_{ws 1}$
modulo the fact that $n_{i} \tilde \sigma^{i}_{ 1}$ is a full one-superform 
in the superspace and $ n_{i} \tilde \sigma^{i}_{ws 1}$ is its pull-back on 
the w.s.

Now we must be more precise about the action of $ \delta $. There are two 
equivalent options ($\delta$ and $\tilde \delta$):

1)   $\delta$ induces Lorentz and gauge transformations with parameters 
$\psi^{ab}$,  $ \sigma^{i}_{1}$ and $\sigma_{3}$.\\ 
 In this case $\delta$ and $s$ anticommute and $ s $ is nilpotent. 
But $ s $ induces also Lorentz and gauge transformations
 with  parameters  $ \lambda^{ \gamma} {\Omega'}_{ \gamma}{}^{ab} +
\lambda^{\ast \gamma} {\Omega'}_{ \gamma}^{\ast ab}$ etc.

2)  $\tilde \delta$ induces Lorentz and gauge transformations with 
parameters 
$\tilde \psi^{ab}$, $\tilde \sigma_{1}$,  and $\tilde \sigma_{3}$.\\
Writing $s + \delta = \tilde s + \tilde \delta$, now
 $\tilde \delta$ and $\tilde s$ do not anticommute and $\tilde s$ is not 
nilpotent.
Now  $\tilde s $ induces covariant transformations
($\tilde s$ is the covariant BRST differential).

The BRST transformations of fields and ghost can be obtained by expanding in
ghost number the parametrizations of the extended curvatures.

In the sector with ghost number $n_{gh}= 0 $ it reproduces the parametrization 
on which we started. In the sector with $n_{gh}= 1$ it gives the BRST transformations 
of fields and forms. In the sector with $n_{gh} = 2 $ it yields the transformations 
of the ghosts (and the ghost constraints as we shall see).

\subsection{\bf Parametrization of Extendended Torsions and Curvatures}

As already noted, it is straightforward to compute the parametrization of
torsions and curvatures with the rescaled and redefined fields and forms. 
Here we will give the results of this computation directly for the extended 
objects.

The extended version of (\ref{29a}) and (\ref{33}) are
\begin{eqnarray}
\hat \Delta \phi =  - (\hat {\cal E}^{\ast}
\pi^{\ast}) + {1 \over 2} {\cal E}^{a}(\rho^{\ast}_{a} - 16 \kappa_{a}),
\label{34}
\end{eqnarray}
\begin{eqnarray}
\hat \Delta {\cal E}^{a} = (\hat {\cal E}^{\ast}\Gamma^{a}\hat {\cal E}),
\label{35}
\end{eqnarray}
where, in (\ref{35}) and in the following, $ \hat \Delta $ denotes the 
extension of 
$\Delta' $, the covariant differential associated to the Lorentz connection 
${\Omega'}^{ab}$, defined in (\ref{32a}) and (\ref{32b}).
 
Then
\begin{eqnarray}
\hat \Delta \pi_{\alpha} = -{1 \over {24}}\hat {\cal E}^{\beta}(\Gamma^{abc})
_{\beta \alpha} ({\cal H}_{abc} -{1 \over 4}  \pi_{abc}) + \hat {\cal E}^{
\ast \beta}[(\Gamma^{c})_{\beta \alpha}(8 \kappa_{c} + {1 \over 2} \rho_{c}) -
 {1 \over 3} (\Gamma^{abc})_{\beta \alpha}\kappa_{abc}] + 
{\cal E}^{c}\Delta_{c}\pi_{\alpha},
\label{36}
\end{eqnarray}
where $\pi_{abc} = (\pi \Gamma_{abc}\pi) $ (and  $\pi^{\ast}_{abc} = 
(\pi^{\ast}\Gamma_{abc}\pi^{\ast}) $ ).

\begin{eqnarray}
\hat \Delta \hat {\cal E}^{\alpha} = {1 \over 2}[ (\hat {\cal E} \Gamma^{c}
\hat {\cal E})(\Gamma_{c}\pi)^{\alpha} +   (\hat {\cal E}^{\ast} \Gamma^{c}
\hat {\cal E}^{\ast})(\Gamma_{c}\pi)^{\alpha}] +  (\hat {\cal E} \Gamma^{c}
\hat {\cal E}^{\ast})(\Gamma_{c}\pi^{\ast})^{\alpha} - 
%\nonumber
%\end{eqnarray}
%\begin{eqnarray}
\hat {\cal E}^{\alpha}[ (\hat {\cal E}\pi) +  (\hat {\cal E}^{\ast}\pi^{
\ast})] 
\nonumber
\end{eqnarray}
\begin{eqnarray}
- \hat {\cal E}^{\ast{}\alpha}[ (\hat {\cal E}^{\ast}\pi) +  
(\hat {\cal E}\pi^{\ast})]
+{1 \over 8}{\cal E}^{c}(\hat {\cal E}^{\ast}\Gamma^{ab})^{\alpha}[({\cal H}_{
abc} + {\cal H}^{\ast}_{abc}) - {3 \over 4} (\pi_{abc} + \pi^{\ast}_{abc})]
%+ {1 over 4} {\cal E}^{c}\hat {\cal E}^{\alpha}(\rho_{c} + \rho^{\ast}_{c})
\nonumber
\end{eqnarray}
\begin{eqnarray}
+ {\cal E}^{c} \hat {\cal E}^{\ast \beta} {1 \over {48}}(\Gamma^{c}
\Gamma^{abd})_{\beta}{}^{\alpha}[ 
({\cal H}_{abd} - {\cal H}^{\ast}_{abd}) - {1 \over 4}( \pi_{abd} -
\pi^{\ast}_{abd} ) ]  
\nonumber
\end{eqnarray}
\begin{eqnarray}
+  {\cal E}^{c} \hat {\cal E}^{\beta}\lbrace {1 \over {48}}(\Gamma^{
abde})_{\beta}{}^{\alpha} {1 \over 4}{\cal F}^{(+)}_{abdec} +
 (\Gamma_{c}\Gamma^{b})_{\beta}^{\alpha}[ 4 \kappa_{b} - {1 \over 8}
(\rho_{b} - \rho^{\ast}_{b})] \rbrace + \frac{1}{2} {\cal E}^{c} {\cal E}^{b}
\tau^\alpha_{bc},          
\label{37}
\end{eqnarray}
\begin{eqnarray}
(n_{i}\hat H^{i}_{3}) \equiv \hat d (n_{i}\hat B^{i}_{2}) = {1 \over 2} 
{\cal E}^{a}
[(\hat {\cal E}\Gamma_{a}\hat {\cal E}) + (\hat {\cal E}^{\ast}\Gamma_{a}
\hat {\cal E}^{\ast})] + {\cal E}^{a} {\cal E}^{b} {\cal E}^{c}[{1 \over 6}
({\cal H}_{abc} + {\cal H}^{\ast}_{abc}) - {1 \over 8}(\pi_{abc} + 
\pi^{\ast}_{abc})],
\label{38}
\end{eqnarray}
\begin{eqnarray}
\hat F_{5} = {1 \over 6}{\cal E}^{a}{\cal E}^{b}{\cal E}^{c}
(\hat {\cal E}^{\ast}\Gamma_{cba}\hat {\cal E}) - {1 \over{12}}{\cal E}^{a}
{\cal E}^{b}{\cal E}^{c}{\cal E}^{d}[(\hat {\cal E}^{\ast}\Gamma_{abcd}\pi^{
\ast}) + (\hat {\cal E}\Gamma_{abcd}\pi)] 
\label{38-2}
\end{eqnarray}
\begin{eqnarray}
+ {1 \over {5!}}
{\cal E}^{a}{\cal E}^{b}{\cal E}^{c}{\cal E}^{d}{\cal E}^{e}[
 {\cal F}^{(+)}_{abcde} - 3 \kappa^{(-)}_{abcde}].
\label{39}
\end{eqnarray} 
Moreover, if we call $\hat R^{ab}$ the extended curvature of the redefined 
Lorentz connection ${\Omega'}^{ab}$ one has
\begin{eqnarray}
\hat R^{ab}= \lbrace - {1 \over 4}[\hat {\cal E}\Gamma_{f} \hat {\cal E}) +
(\hat {\cal E}^{\ast}\Gamma_{f} \hat {\cal E}^{\ast})]
 [ ({\cal H}^{fab} + {\cal H}^{\ast fab}) - {3 \over 4}(\pi^{fab}  + \pi^{
\ast fab})]
- {1 \over {48}} [( \hat {\cal E}\Gamma^{a}\Gamma^{fgh}\Gamma^{b}\hat {\cal E})
\nonumber
\end{eqnarray}
\begin{eqnarray}
 -( \hat {\cal E}^{\ast}\Gamma^{a}\Gamma^{fgh}\Gamma^{b}\hat {\cal E}^{\ast})]
 [({\cal H}_{fgh}- {\cal H}^{\ast}_{fgh})
- {1 \over 4} (\pi_{fgh} -\pi^{\ast}_{fgh})] 
+ {1 \over {16 \cdot 5!}}(\hat {\cal E} \Gamma^{a}\Gamma_{fghlm}\Gamma^{b}
\hat {\cal E}^{\ast}) {\cal F}^{(+) fghlm}
\nonumber
\end{eqnarray}
\begin{eqnarray}
+ (\hat {\cal E}\Gamma^{a}
\Gamma^{f}\Gamma^{b}\hat {\cal E}^{\ast})[ 4 \kappa_{f} - {1 \over 8} (
\rho_{f} - \rho^{\ast}_{f})] 
+ c.c.\rbrace - {1 \over 2} {\cal E}^{c}[ 
(\hat {\cal E}^{\ast}\vartheta_{c}^{ab}) +  (\hat {\cal E}
 \vartheta^{\ast ab }_{c})] + {1 \over 2} {\cal E}^{c}{\cal E}^{d}{\cal R}_{
dc}{}^{ab},  
\label{40}
\end{eqnarray}
where the explicit forms of  $ \vartheta_{c}^{ab} $ and 
$ {\cal R}_{dc}{}^{ab}$ in terms of the other fields are not needed.
Notice that $\hat F_{5}$ is purely imaginary while $\hat R^{ab}$, $\hat H_{3}$ and
 $\hat \Delta {\cal E}^{a}$ are real, and the parametrizations of $ \hat \Delta
\phi^{\ast}$, $\hat \Delta \pi$ and $\hat \Delta \hat {\cal E}^{\ast {}
\alpha}$ can be obtained by taking the complex conjugate of (\ref{34}),
(\ref{36}) and (\ref{37}), respectively.

\subsection{\bf Pure spinor constraints}

As already mentioned, equations (\ref{34}) - (\ref{40}) at ghost number $n_{gh}
= 0 $ give the parametrizations of the torsions and curvatures of the rescaled 
and redefined fields and forms and at ghost number $n_{gh}= 1 $ they give the
BRST transformations of these fields and forms. Now we are interested in the 
sector with ghost number $n_{gh} = 2$ where these equations give the BRST
transformations of the ghosts. The vanishing of $\lambda^{a}$, i.e.,
  Eq. (\ref{336}) together 
with (\ref{35}) at $n_{gh} = 2 $ yields the constraint
\begin{eqnarray}
(\lambda \Gamma^{a}\lambda^{\ast}) = 0,
\label{41}
\end{eqnarray}
and the vanishing of $ n_{i}\tilde \sigma_{1}^{i} $, i.e., equation (\ref{337})
together with (\ref{38}) implies $ {\cal E}^{a}[(\lambda\Gamma_{a}\lambda) +
 (\lambda^{\ast}\Gamma_{a}\lambda^{\ast})] = 0 $ so that 
\begin{eqnarray}
(\lambda \Gamma^{a}\lambda) + (\lambda^{\ast}\Gamma^{a}\lambda^{\ast}) = 0.
\label{42}
\end{eqnarray}
However, it is fair to recall, as remarked before, that equation (\ref{337}) is
stronger than the condition  $ n_{i}\tilde \sigma_{1 ws}^{i} = 0 $ that 
follows from the requirement that the pull-back of the G-S lagrangian vanishes 
along $\kappa$. Indeed   $ n_{i}\tilde \sigma_{1 ws}^{i} = 0 $ only implies 
that the pull-back $ {\cal E}_{\pm}^{a}[(\lambda\Gamma_{a}\lambda) +
 (\lambda^{\ast}\Gamma_{a}\lambda^{\ast})] $ vanishes.

The constraints (\ref{41}), (\ref{42}) gain a more standard form if one 
writes
 $\lambda^{\alpha}= {1 \over {\sqrt 2}}( \lambda_{1}^{\alpha}+ i 
\lambda_{2}^{\alpha}) $. In fact, in terms of $\lambda_{1}^{\alpha}$ and 
 $\lambda_{2}^{\alpha}$ these constraints become 
\begin{eqnarray}
(\lambda_{1} \Gamma^{a}\lambda_{1}) = 0 = (\lambda_{2}\Gamma^{a}
\lambda_{2}),
\label{43}
\end{eqnarray}
i.e., the pure spinor constraints for $\lambda_{1}$ and $\lambda_{2}$.
Moreover, (\ref{37}) in the sector $n_{gh} = 2$ gives the BRST transformation
of $\lambda^{\alpha}$
\begin{eqnarray}
\tilde s \lambda^{\alpha} = - \lambda^{\alpha}[(\lambda\pi) + (\lambda^{\ast}
\pi^{\ast})] -  \lambda^{\ast \alpha}[(\lambda^{\ast}\pi) + (\lambda
\pi^{\ast})],
\label{43b}
\end{eqnarray}
or in terms of $\lambda_{1}$ and $\lambda_{2}$,
\begin{eqnarray}
 \tilde s \lambda_{1}^{\alpha} = - 2 \lambda_{1}^{\alpha}(\lambda_{1}\pi_{1}),
\qquad \tilde s \lambda_{2}^{\alpha} = 2 \lambda_{2}^{\alpha}(\lambda_{2}
\pi_{2}).
\label{43c}
\end{eqnarray}
\subsection{\bf Real and Imaginary Components of Spinor-like Vielbeins and 
Other Fields}
These results suggest that it should be convenient to rewrite equations
(\ref{35}), (\ref{37}), (\ref{38}), (\ref{39}) and (\ref{40}) in terms of the real
 and imaginary components of the 
relevant fields by writing
\begin{eqnarray}
 {\cal E}^{\alpha} = {1 \over {\sqrt 2}}( {\cal E}_{1}^{\alpha} + i
{\cal E}_{2}^{\alpha} ),
\nonumber
\end{eqnarray}
\begin{eqnarray}
\pi^{\alpha} = {1 \over {\sqrt 2}}( \pi_{1}^{\alpha} + i \pi_{2}^{\alpha} ),
\label{44}
\end{eqnarray}
but we will define ${\cal H}_{abc} = {\cal H}_{1 abc} + i {\cal H}_{2 abc}$ and
$ \rho_{a} = \rho_{1 a} + i \rho_{2 a}$. 
Then equations (\ref{35}) - (\ref{40}) yield
\begin{eqnarray}
\hat \Delta {\cal E}^{a} = {1 \over 2}[ (\hat {\cal E}_{1}\Gamma^{a}\hat {
\cal E}_{1}) + (\hat {\cal E}_{2}\Gamma^{a}\hat {\cal E}_{2})],
\label{35a}
\end{eqnarray}
\begin{eqnarray}
\hat \Delta \hat {\cal E}_{1}^{\alpha} = (\hat {\cal E}_{1}\Gamma^{c}
\hat {\cal E}_{1})(\Gamma_{c}\pi_{1})^{\alpha} - 2 \hat {\cal E}_{1}^{\alpha}(
\hat {\cal E}_{1}\pi_{1}) +{1 \over 4} {\cal E}^{c} \hat {\cal E}_{1}^{
\beta}(\Gamma_{ab})_{\beta}{}^{\alpha} {\tilde {\cal H}}_{abc}
\nonumber
\end{eqnarray}
\begin{eqnarray}
- {\cal E}^{c}(M \Gamma_{c}\hat {\cal E}_{2})^{\alpha}
  + {1 \over 2} {\cal E}^{a}{\cal E}^{b}
\tau^\alpha_{1, ba}, 
\label{37a}
\end{eqnarray}
\begin{eqnarray}
\hat \Delta \hat {\cal E}_{2}^{\alpha} = - [(\hat {\cal E}_{2}\Gamma^{c}\hat {
\cal E}_{2})(\Gamma_{c}\pi_{2})^{\alpha} - 2 \hat {\cal E}_{2}^{\alpha}(
\hat {\cal E}_{2}\pi_{2})] + {1 \over 4} {\cal E}^{c} \hat {\cal E}_{2}^{\beta}
(\Gamma_{ab})_{\beta}{}^{\alpha} {\tilde {\cal H}}_{ abc}] 
\nonumber
\end{eqnarray}
\begin{eqnarray}
+ {\cal E}^{c}(\hat {\cal E}_{1}\Gamma_{c} M)^{\alpha}
 + {1 \over 2} {\cal E}^{a}{\cal E}^{b}
\tau^\alpha_{2, ba}, 
\label{37b}
\end{eqnarray}
where we have defined
\begin{eqnarray}
{\tilde {\cal H}}_{abc} = {\cal H}_{1 abc} - {3 \over 4}[ ( \pi_{1}\Gamma_{
abc}\pi_{1}) - (\pi_{2}\Gamma_{abc}\pi_{2})],
\label{37d}
\end{eqnarray}
and
\begin{eqnarray}
M^{\beta \alpha} = (\Gamma^{b})^{\beta \alpha}
({1 \over 4}\rho_{2 b} + 4 \bar \kappa^{b}) +
{1 \over {16 \cdot 5!}}(\Gamma^{abcde})^{\beta \alpha}
  \bar {\cal F}^{(+)}_{abcde}  
\nonumber
\end{eqnarray}
\begin{eqnarray}
+ {1 \over {24}}(\Gamma^{bcd})^{\beta \alpha}[{\cal H}_{2 bcd} - {1 \over 4}
(\pi_{1}\Gamma_{abc}\pi_{2})],
\label{37c}
\end{eqnarray}
and we have written $ \bar \kappa_{a} = i \kappa_{a}$, 
$ \bar \kappa^{(-)}_{abcde} = i \kappa^{(-)}_{abcde}$ and
$\bar {\cal F}^{(+)}_{abcde} = i {\cal F}^{(+)}_{abcde}$ so that
$ \bar \kappa_{a}$, $ \bar \kappa^{(-)}_{abcde} $ and
$\bar {\cal F}^{(+)}_{abcde}$ are real. The expression of the fields $
\tau_{1/2,ab}$ is irrelevant for our purposes.   Moreover
\begin{eqnarray}
(n_{i}\hat H^{i}_{3}) \equiv \hat d (n_{i}\hat B^{i}_{2}) = {1 \over 2}
 {\cal E}^{a}
[(\hat {\cal E}_{1}\Gamma_{a}\hat {\cal E}_{1}) - (\hat {\cal E}_{2}\Gamma_{a}
\hat {\cal E}_{2})] + {1 \over 3} {\cal E}^{a} {\cal E}^{b} {\cal E}^{c}
{\tilde {\cal H}}_{abc}, 
\label{38a}
\end{eqnarray}
\begin{eqnarray}
\hat F_{5} = {i \over 6}\lbrace {\cal E}^{a}{\cal E}^{b}{\cal E}^{c}
( \hat {\cal E}_{1}\Gamma_{cba}\hat {\cal E}_{2}) - {1 \over 2} {\cal E}^{a}
{\cal E}^{b}{\cal E}^{c}{\cal E}^{d}(\hat {\cal E}_{2}\Gamma_{abcd}\pi_{2})
  + {1 \over {20}}
{\cal E}^{a}{\cal E}^{b}{\cal E}^{c}{\cal E}^{d}{\cal E}^{e}[
\bar {\cal F}^{(+)}_{abcde} - 3 \bar \kappa^{(-)}_{abcde}] \rbrace,
\label{39a}
\end{eqnarray} 
and
\begin{eqnarray}
\hat R^{ab} = - {1 \over 4} [ (\hat {\cal E}_{1}\Gamma_ {f}\hat {\cal E}_{1}) 
- (\hat {\cal E}_{2}\Gamma_ {f}\hat {\cal E}_{2})] \tilde {\cal H}^{fab} 
+ 2 (\hat {\cal E}_{1} \Gamma^{a}M\Gamma^{b}
\hat {\cal E}_{2})  - {1 \over 2} {\cal E}^{c}[ 
(\hat {\cal E}_{1}\vartheta_{1 c}^{ab}) +  (\hat {\cal E}_{2}
 \vartheta^{ ab }_{2 c})] 
\nonumber
\end{eqnarray}
\begin{eqnarray}
+ {1 \over 2} {\cal E}^{c}{\cal E}^{d}{\cal R}_{
dc}{}^{ab}.  
\label{40a}
\end{eqnarray}
Using the identity  $$   ({\cal E}_{1}\Gamma^{a}{\cal E}_{1})
(\Gamma_{a}\pi_{1})^{\alpha} - 2 {\cal E}_{1}^{\alpha}({\cal E}_{1} \pi_{1}) 
= - {1 \over 4}[ 
 (\Gamma^{ab}{\cal E}_{1})^{ \alpha}({\cal E}_{1}\Gamma_{ab}\pi_{1}) - 2
{\cal E}_{1}^{\alpha} ({\cal E}_{1}\pi_{1})], $$ 
and a similar one for ${\cal E}_{2}$, the sectors (0,2) of $\hat \Delta 
\hat {\cal E}_{i}$  
can be rewritten as
\begin{eqnarray}
(\hat \Delta \hat {\cal E}_{1}^{\alpha})_{(0,2)} = - {1 \over 4}[(\Gamma^{ab}
\hat {\cal E}_{1})^{\alpha}(\hat {\cal E}_{1}\Gamma_{ab}\pi_{1}) - 2
\hat {\cal E}_{1}^{\alpha} (\hat {\cal E}_{1}\pi_{1})],
\label{41a}
\end{eqnarray}
\begin{eqnarray}
(\hat \Delta \hat {\cal E}_{2}^{\alpha})_{(0, 2)} = {1 \over 4}[(\Gamma^{ab}
\hat {\cal E}_{2})^{\alpha}(\hat {\cal E}_{2}\Gamma_{ab}\pi_{2}) - 2
\hat {\cal E}_{2}^{\alpha} (\hat {\cal E}_{2} \pi_{2})]. 
\label{41b}
\end{eqnarray}
Moreover
\begin{eqnarray}
\hat \Delta  \pi_{1 \alpha} = \hat {\cal E}_{1}^{\beta}[ - {1 \over {24}} 
(\Gamma^{abc})_{\beta \alpha} \tilde{\cal H}_{abc} +
{1 \over 2} (\Gamma^{c})_{\beta \alpha}\rho_{1 c} - {1 \over 48} (\Gamma^{abc}
)_{\beta \alpha} (\pi_{1}\Gamma_{abc}\pi_{1})]
\nonumber
\end{eqnarray}
\begin{eqnarray}
- {1 \over 4}(\Gamma^{c}M\Gamma_{c}\hat {\cal E}_{2})_{\alpha}  + 
{\cal E}^{c}\hat \Delta \pi_{ 1 \alpha},
\label{41c}
\end{eqnarray}
\begin{eqnarray}
\hat \Delta  \pi_{2 \alpha} = \hat {\cal E}_{2}^{\beta}[ - {1 \over {24}} 
(\Gamma^{abc})_{\beta \alpha} \tilde {\cal H}_{abc} +
{1 \over 2} (\Gamma^{c})_{\beta \alpha}\rho_{1 c} - {1 \over 48} 
(\Gamma^{abc})_{\beta \alpha} (\pi_{2}\Gamma_{abc}\pi_{2})] 
\nonumber
\end{eqnarray}
\begin{eqnarray}
+ {1 \over 4}(\hat {\cal E}_{1}
\Gamma^{c} M \Gamma_{c})_{\alpha} + {\cal E}^{c}\hat \Delta \pi_{ 2 \alpha}.
\nonumber
\end{eqnarray}
\begin{eqnarray}
\label{41d}
\end{eqnarray}

The parametrizations of torsions and curvatures are obtained by looking at the 
sector with $n_{gh}= 0 $ of equations (\ref{44}) - (\ref{41d}), i.e., 
dropping the hats in these equations.

As for the sector with $n_{gh} = 1 $, let us report only the BRST 
transformations of the supervielbeins ${\cal E}^{a}$ , ${\cal E}_{i}^{\alpha}
 $ and the B-fields $ n_{i}B^{i}_{2}$:
\begin{eqnarray}
\tilde s {\cal E}^{a} =  (\lambda_{1}\Gamma^{a}{\cal E}_{1}) +   
  (\lambda_{2}\Gamma^{a}{\cal E}_{2}),
\label{101}
\end{eqnarray}
\begin{eqnarray}
\tilde s {\cal E}_{1}^{\alpha} = - \Delta \lambda_{1}^{\alpha}  -
{1 \over 4}(\Gamma^{ab}\lambda_{1})^{\alpha}( {\cal E}_{1}\Gamma_{ab}\pi_{1})
+ {1 \over 2} \lambda_{1}^{\alpha}({\cal E}_{1}\pi_{1}) 
- {1 \over 4} {\cal E}^{c} (\Gamma_{ab} \lambda_{1})^{\alpha}\tilde{\cal H}_{ 
abc}
\nonumber
\end{eqnarray}
\begin{eqnarray}
 - {1 \over 4}(\Gamma^{ab}{\cal E}_{1})^{\alpha}( \lambda_{1}\Gamma_{ab}
\pi_{1}) + {1 \over 2}  {\cal E}_{1}^{\alpha} (
\lambda_{1}\pi_{1}) - {\cal E}^{c}(M \Gamma_{c}\lambda_{2})^{\alpha},
\label{102}
\end{eqnarray}
\begin{eqnarray}
\tilde s {\cal E}_{2}^{\alpha} = - \Delta \lambda_{2}^{\alpha}  +
{1 \over 4}(\Gamma^{ab}\lambda_{2})^{\alpha}( {\cal E}_{2}\Gamma_{ab}\pi_{2})
 - {1 \over 2} \lambda_{2}^{\alpha}({\cal E}_{2}\pi_{2})] 
+ {1 \over 4} {\cal E}^{c} (\Gamma_{ab} \lambda_{2})^{\alpha}\tilde {\cal H}_{
 abc} 
\nonumber
\end{eqnarray}
\begin{eqnarray}
+{1 \over 4}(\Gamma^{ab}{\cal E}_{2})^{
\alpha}( \lambda_{2}\Gamma_{ab}\pi_{2}) - {1 \over 2}{\cal E}_{2}^{\alpha} (
\lambda_{2}\pi_{2})] + {\cal E}^{c}(\lambda_{1}\Gamma_{c} M)^{\alpha},
\label{103}
\end{eqnarray}
 \begin{eqnarray}
s (n_{i}B^{i}_{2}) =    {\cal E}^{a}
[(\lambda_{1}\Gamma_{a} {\cal E}_{1}) - (\lambda_{2}\Gamma_{a}
{\cal E}_{2})]. 
 \label{104}
\end{eqnarray}
In this notation, the Green-Schwarz action (\ref{16}) is
\begin{eqnarray}
I_{GS} = {1 \over 2} \int [ {\cal E}_{+}^{a}{\cal E}_{- a } + 
2 n_{i}B_{2}^{i}],
\label{105}
\end{eqnarray}
where $ {\cal E}_{\pm}^{A} $ are the left-handed and right-handed pullbacks 
of the supervielbeins on the worldsheet.
The BRST transformation of $I_{GS}$ is
\begin{eqnarray}
 sI_{GS} =   \int  [(\lambda_{1}\Gamma_{a}
{\cal E}_{+}^{a}{\cal E}_{- 1}) + (\lambda_{2}\Gamma_{a}{\cal E}_{-}^a{\cal E}_{+ 2})].
\label{106}
\end{eqnarray}
Notice that, if one chooses $ \lambda_{1}^{\alpha} = (k_{1} \Gamma^{b}{
\cal E}_{+ b})^{\alpha}$ and $   \lambda_{2}^{\alpha} = (k_{2} \Gamma^{b}{
\cal E}_{- b})^{\alpha}$ where $k_{i}$ are local parameters,  $I_{GS}$ is 
invariant (modulo the Virasoro constraints).
This is the $\kappa$-symmetry of the G-S action.

A useful identity that follows from the Bianchi identity 
$\hat \Delta \hat R^{ab} = 0$ in the sector with ghost number $3$, is
\begin{eqnarray}
(\lambda_{1}\Gamma^{[a}[\lambda_{1}^{\alpha}\Delta_{1 \alpha} P +
\lambda_{2}^{\alpha}\Delta_{2 \alpha} P]\Gamma^{b]}\lambda_{2}) = 0,
\label{45a}
\end{eqnarray}
where $$ M = e^{-2(\phi + \phi^{*})} P .$$
If one defines 
\begin{eqnarray}
\lambda_{1}{}^{\alpha}\Delta_{1 \alpha} P^{\beta \gamma} = \lambda_{1}{}^{
\alpha} C_{1 \alpha}{}^{\beta \gamma},
\nonumber
\end{eqnarray}
\begin{eqnarray}
\lambda_{2}{}^{\alpha}\Delta_{2 \alpha} P^{\beta \gamma} = \lambda_{2}{}^{
\alpha}C_{2 \alpha}{}^{\beta \gamma},
\label{45b}
\end{eqnarray}
(\ref{45a}) implies that $C_{1 \alpha}{}^{\beta \gamma}$ and   
 $C_{2 \alpha}{}^{\beta \gamma}$ are Lorentz-Weyl valued in $\alpha$, $\beta $
 and $ \alpha $, $\gamma$ respectively.

It is interesting to note that the parametrization of torsions and 
curvatures, when expressed in terms of the real and imaginary parts of the 
rescaled and redefined fields and forms, has a structure very similar to that
  of IIA superstring \cite{Ton} , a non-surprising result given
 that it is also present from the beginning in the treatment of Berkovits and
Howe in \cite{BH}.

\section {\bf  Pure Spinor Action}

\subsection{\bf Antighosts, $d_{\underline{\alpha}}$ Fields and 
BRST charge} 
In order to derive the pure spinor action one must add to the 
superspace coordinates $Z^{M} = (X^a, \theta^\mu)$ the ghosts $$\lambda^{\underline{\alpha}} 
= (\lambda_{1}^{\alpha}, \lambda_{2}^{ \alpha}), $$ the antighosts $$ 
\omega_{
\underline{\alpha}} = (\omega_{1 \alpha}, \omega_{2  \alpha}),$$ with ghost 
number $n_{gh} = - 1 $ 
which will play the role of the conjugate momenta of $ \lambda^{\underline{\alpha}}$, 
and the fields $$ d_{\underline{\alpha}} = ( d_{1 \alpha}, d_{2  \alpha}),$$
that will also play the role of the conjugate momenta of $\theta^\mu$ and are essentially the
BRST partners of $\omega_{\underline{\alpha}}$.
 From the worldsheet point of view,
$\omega_{1}$, $d_{1}$ and $\omega_{2}$, $d_{2}$ are left-handed and
right-handed chiral  fields, respectively.

An index $\underline{ \alpha}$
 repeated, like for instance  in $\lambda^{\underline{\alpha}}d_{\underline{
\alpha}}$,
means $ \lambda^{\underline{\alpha}}d_{\underline{\alpha}} = (\lambda_{1}
d_{1})+  (\lambda_{2}d_{2})$ whereas indices $i (i= 1,2)$ repeated
do not imply summation.

Since, as a consequence of the pure spinor constraints, $\lambda^{
\underline{\alpha}}$ contains 11  + 11 degrees of freedom, also $\omega_{
\underline{\alpha}}$ should contain 11 + 11 independent components. This is
realized by assuming that the pure spinor action is invariant under the
$\omega$-gauge symmetry
%%%%%% 4.10%%%%%%%%%%%%%%%%
\begin{eqnarray}
 \delta^{(\omega)} \omega_{i} = \Lambda_{i }^{a}(\Gamma_{a}\lambda_{i}) 
\qquad i = 1, 2,
\label{47}
\end{eqnarray}
%%%%%%%%%%%%%%%%%%%%%%%%%%%
 where $\Lambda_{i}^{a}$ are local gauge parameters.
The  $d_{\underline{\alpha}}$
allow us to define the  BRST charge 
\begin{eqnarray}
 Q = \oint (\lambda^{\underline{
\alpha}}d_{\underline{\alpha}}) = \oint (\lambda_{1} d_{1})  + 
\oint (\lambda_{2} d_{2}), 
\label{48}
\end{eqnarray}
 that generates the 
transformations induced by the BRST differential $s$.
It is also useful to split $s$ as $ s = s_{1} + s_{2}$ where $s_{1}$ is 
generated by the charge $Q_{1} = \oint (\lambda_{1} d_{1})$ and  $s_{2}$ is 
generated by the charge $Q_{2} = \oint (\lambda_{2} d_{2})$.

In order to specify $Q$,  prove its nilpotence and compute the BRST 
transformations of $\lambda_{i}$,
$\omega_{i}$ and $d_{i}$ one needs the expression of $(\lambda_{i}d_{i})$
which is expected to be
\begin{eqnarray}
\lambda_{1}^{\alpha}d_{1 \alpha} = \lambda_{1}^{\alpha}[d^{(0)}_{1 \alpha} +
 (\Omega'_{ \alpha \beta}{}^{\gamma} +
X^{(1)}_{ \alpha \beta}{}^{\gamma}) \omega_{1 \gamma}\lambda_{1}^{\beta} +
 \Omega'_{\alpha  \beta}{}^{\gamma} \omega_{2 \gamma}
\lambda_{2}^{ \beta}],
 \label{50}
\end{eqnarray}
\begin{eqnarray}
 \lambda_{2}^{\alpha} d_{2 \alpha} =\lambda_{2}^{ \alpha}[ 
d^{(0)}_{2  \alpha} +
 (\Omega'_{\alpha  \beta}{}^{ \gamma} +
X^{(2)}_{ \alpha  \beta}{}^{ \gamma}) \omega_{2 \gamma}
\lambda_{2}^{\beta} +
 \Omega'_{ \alpha \beta}{}^{\gamma}  \omega_{1 \gamma}\lambda_{1}^{\beta}],
 \label{50bis}
\end{eqnarray}
where $ d^{(0)}_{\underline{\alpha}}$, acting on superfields, induces the
 tangent space derivative $D_{\underline{\alpha}}$ and $ \Omega'_{
\underline{\alpha}}{}^{\underline{\beta}}$ are the spinorial partners of the 
Lorentz connection defined in (\ref{32a}) and (\ref{32b}).  The 
superfields $ X^{(1)}$ and $X^{(2)}$ are needed to assure the nilpotence of 
$ Q $ and to  reproduce equation  (\ref{43c}).

As we will see, these two requirements are satisfied if one chooses for 
$ X^{(i)}$
\begin{eqnarray}
 X^{(1)}_{\alpha \beta}{}^{\gamma} = - {1 \over 4} (\Gamma^{ab}\pi_{1})_{
\alpha}(\Gamma_{ab})_{\beta}{}^{\gamma} - {1 \over 2} \pi_{1 \alpha}\delta_{
\beta}{}^{\gamma},
\label{53a}
\end{eqnarray}
\begin{eqnarray}
 X^{(2)}_{ \alpha  \beta}{}^{ \gamma} =  {1 \over 4} (\Gamma^{ab}
\pi_{2})_{ \alpha}(\Gamma_{ab})_{ \beta}{}^{ \gamma} + {1 \over 2} \pi_{2  
\alpha}\delta_{ \beta}{}^{ \gamma}].
\label{53b}
\end{eqnarray}
We shall write
$$ X_{\underline{\alpha} \underline{\beta}}{}^{\underline{\gamma}} = (
 X^{(1)}_{\alpha \beta}{}^{\gamma}, X^{(2)}_{\alpha  \beta}{}^{
\gamma}),$$ 
and also define the Lorentz-Weyl connections $ \tilde \Omega_{ 
\underline{\beta}}{}^{\underline{\gamma}} =\Omega'_{ \underline{\beta}}{}^{
\underline{\gamma}} +  X_{ \underline{\beta}}{}^{\underline{\gamma}}$
with being $$  X_{ \underline{\beta}}{}^{\underline{\gamma}} = 
 {\cal E}^{\underline{\alpha}}X_{\underline{\alpha} \underline{\beta}}{}^{
\underline{\gamma}} + {\cal E}^{c} {\cal H}_{c \underline{\beta}}{}^{
\underline{\gamma}}. $$  It is convenient to choose
 $$  {\cal H}_{c \underline{\beta}}{}^{
\underline{\gamma}} =  {1 \over 4} (\Gamma^{ab})_{\underline{\beta}}{}^{
\underline{\gamma}}\tilde {\cal H}_{ abc},   $$ so that, calling
 $\tilde \Delta $  the covariant differential related to the 
Lorentz-Weyl connection  $\tilde \Omega $, the torsion
 $ \hat{\tilde T}^{
\underline{\alpha}} \equiv \hat{ \tilde \Delta}\hat {\cal E}^{\underline{
\alpha}}
 $ vanishes in the sector (0,2), and in the sector (1, 1) it becomes
\begin{eqnarray}
(\hat {\tilde T}^{\underline{\alpha}})_{(1, 1)} = ( - {\cal E}^{c}(M \Gamma_{c}
\hat {\cal E}_{2})^{\alpha}, \quad {\cal E}^{c}(\hat {\cal E}_{1}\Gamma_{c}
M)^{\hat \alpha} ).
\label{54a}
\end{eqnarray}
Then, from the Bianchi 
identity $ \tilde \Delta \tilde \Delta {\cal E}^{\underline{\alpha}} = 
{\cal E}^{\underline{\beta}} \tilde R_{\underline{\beta}}{}^{
\underline{\alpha}}$  one obtains for $ {\cal E}^{\underline{\beta}}
\tilde R_{\underline{\beta}}{}^{\underline{\alpha}}$ the simple result
\begin{eqnarray}
({\cal E}_{1}\tilde R^{(1) \gamma})_{(0,3)} = - {1 \over 2} ({\cal E}_{1}
\Gamma^{a}{\cal E}_{1})(M\Gamma_{a}{\cal E}_{2})^{\gamma},
\label{64}
\end{eqnarray}
\begin{eqnarray}
({\cal E}_{2 \beta}\tilde R_{\beta}^{(2) \gamma})_{(0,3)}= {1 \over 2}
 ({\cal E}_{2}\Gamma^{a}{\cal E}_{2})({\cal E}_{1}\Gamma_{a} M)^{ \gamma}.
\label{65}
\end{eqnarray}

Now, in this notation,  
\begin{eqnarray}
d_{\underline{\alpha}} = d^{(0)}_{\underline{\alpha}} + 
 \tilde \Omega_{\underline{\alpha} \underline{\beta}}{}^{\underline{\gamma}}
\lambda^{\underline{\beta}}\omega_{\underline{\gamma}},
\label{54}
\end{eqnarray}
and  
\begin{eqnarray}
Q = \oint \lambda^{\underline{\alpha}}(d^{(0)}_{\underline{\alpha}} + 
 \tilde \Omega_{\underline{\alpha} \underline{\beta}}{}^{\underline{\gamma}}
\lambda^{\underline{\beta}}\omega_{\underline{\gamma}}).
\label{55}
\end{eqnarray}
Moreover the BRST transformations of $\lambda_{i}$ and $\omega_{i}$  become
\begin{eqnarray}
s \lambda^{\underline{\alpha}} = \lambda^{\underline{\beta}}\lambda^{
\underline{\gamma}}
\tilde \Omega_{\underline{\beta} \underline{\gamma}}{}^{\underline{\alpha}}, 
\label{56}
\end{eqnarray}
 and
\begin{eqnarray}
s \omega_{\underline{\alpha}} = - d_{\underline{\alpha}} - \lambda^{
\underline{\beta}} \tilde \Omega_{\underline{\beta} \underline{\alpha}}{}^{
\underline{\gamma}}\omega_{\underline{\gamma}}.
\label{57}
\end{eqnarray}
It is useful to notice that if one defines
\begin{eqnarray}
Y^{(1)}_{\alpha\beta}{}^{\gamma} =  X^{(1)}_{ \alpha  \beta}{}^{ \gamma}
 +  2 \pi_{1 \alpha}\delta_{ \beta}{}^{ \gamma}, \qquad 
Y^{(2)}_{\alpha\beta}{}^{\gamma} =  X^{(2)}_{ \alpha  \beta}{}^{ \gamma}
 -  2 \pi_{2 \alpha}\delta_{ \beta}{}^{ \gamma},
\label{53c}
\end{eqnarray}
$ Y^{(i)}_{\alpha\beta}{}^{\gamma}$ are symmetric in $ \alpha$ and $\beta$ 
and therefore Lorentz-Weyl valued both in $\alpha , \gamma$ and in $
\beta, \gamma$.
Given (\ref{53a}) and (\ref{53b}), the covariant form of (\ref{56}), that is,
$ \tilde s \lambda_{i}^{\alpha} =  \lambda_{i}^{\beta}\lambda_{i}^{\gamma}
X^{(i)}_{\beta \gamma}{}^{\alpha} $ reproduces (\ref{43c}) and using 
(\ref{53c})
\begin{eqnarray}
\tilde s \omega_{i \alpha} = - d_{i \alpha} - \lambda_{i}^{\beta}Y^{(i)}_{
\alpha\beta}{}^{\gamma}\omega_{i \gamma} \mp 2 (\lambda_{i}\pi_{i}) \omega_{
i\alpha}.
\label{57a}
\end{eqnarray} 

To compute $ s d_{\underline{\alpha}}$ we assume
\begin{eqnarray}
\lbrace \oint\lambda^{\underline{\beta}} d_{\underline{\beta}}^{(0)}, d_{
\underline{\alpha}}^{(0)}\rbrace = -(\Gamma^{a}\lambda)_{\underline{\alpha}}[ 
{\cal E}_{\pm a} +   \tilde \Omega_{a}],
\label{58}
\end{eqnarray}
where $$ {\cal E}_{\pm a}\lambda^{\underline{\beta}} \Gamma^{a}_{
\underline{\beta}\underline{\alpha}} = ( {\cal E}_{+}^{a}(\lambda_{1} 
\Gamma_{a})_{ \alpha}, \quad {\cal E}_{-}^{a}(\lambda_{2}
\Gamma_{a})_{ \alpha}), $$
Then
\begin{eqnarray}
s d_{\underline{\alpha}} =  - (\Gamma^{a}\lambda)_{\underline{\alpha}} 
{\cal E}_{\pm a}  + \lambda^{\underline{
\beta}} \tilde \Omega_{\underline{\alpha} \underline{\beta}}{}^{
\underline{\gamma}} d_{\underline{\gamma}} +  \lambda^{\underline{\delta}}
\lambda^{\underline{\beta}}\tilde R_{\underline{\delta} \underline{\alpha}
 \underline{\beta}}{}^{\underline{\gamma}}\omega_{\underline{\gamma}}, 
\label{59}
\end{eqnarray}
so that
\begin{eqnarray}
Q^2 =  \oint \lambda^{\underline{\alpha}}\lambda^{\underline {\beta}}
\lambda^{\underline{\delta}} \tilde R_{\underline{\alpha} \underline{\beta}
\underline{\delta}}{}^{\underline{\gamma}}\omega_{\underline{\gamma}}.
\label{60}
\end{eqnarray}
Then, it follows from (\ref{64}) and (\ref{65}) that indeed $Q^2 =0$. 

>From (\ref{56}) (or (\ref{43c})) it also follows that $$ s^2 \lambda_{
\underline{\alpha}} = 0, $$ in  agreement with the nilpotence of $Q$.

However  $s^{2} \omega_ {\underline{\alpha}} $ does not vanish  
since from (\ref{57}) and (\ref{59}) one has \footnote {In our 
notations $
X_{(\alpha\beta)} = {1 \over 2}(X_{\alpha \beta}+ X_{\beta \alpha})$ and
$X_{[\alpha\beta]} = {1 \over 2}(X_{\alpha \beta} - X_{\beta \alpha})$.}
 $$ s^{2} \omega_{\underline{\alpha}} = (\Gamma_{a}\lambda)_{
\underline{\alpha}}{\cal E}_{\pm a} - 2   \lambda^{\underline{\delta}}
\lambda^{\underline{\beta}}\tilde R_{\underline{\delta} (\underline{\alpha}
 \underline{\beta})}{}^{\underline{\gamma}}\omega_{\underline{\gamma}}, $$ 
that is, using (\ref{64}) and (\ref{65}), 
\begin{eqnarray}
s^{2} \omega_{1 \alpha} = (\Gamma_{a}\lambda_{1})_{\alpha}
[{\cal E}_{+}^{a} +  (\omega_{1}M \Gamma_{a} \lambda_{2})],
\label{63a}
\end{eqnarray}
\begin{eqnarray}
s^{2} \omega_{2 \alpha} = (\Gamma_{a}\lambda_{2})_{\hat \alpha}
[{\cal E}_{-}^{a} -  (\lambda_{1} \Gamma_{a} M \omega_{2})].
\label{63b}
\end{eqnarray}

%%%%%%%%%%%%%%%%%%%%%%%%%%%%
Also $ s^2 d_{\underline{\alpha}}$ does not vanish since
\begin{eqnarray}
s^2 d_{1 \alpha} =  - [({\cal E}_{2 +}\Gamma_{a}\lambda_{2})(\lambda_{1}
\Gamma^{a})_{\alpha} + s[ (\omega_{1}M \Gamma_{a}\lambda_{2})(\lambda_{1}
\Gamma^{a})_{\alpha}],
\label{64a}
\end{eqnarray}
\begin{eqnarray}
s^2 d_{2 \alpha} = - [(\Gamma^{a}\lambda_{2})_{ \alpha} (\lambda_{1}
\Gamma_{a}{\cal E}_{1 -})- s [(\Gamma^{a}\lambda_{2})_{\alpha}(\lambda_{1}
\Gamma_{a} M \omega_{2})].
\label{64b}
\end{eqnarray}
The nonvanishing of (\ref{64a}) and (\ref{64b}) is not a problem 
since, as we will see later, the RHS of (\ref{64a}) and (\ref{64b}),
vanishes on shell, being proportional to the fields equations of 
$d_{\underline{\alpha}}$
\begin{eqnarray}
(\lambda_{1}\Gamma^{a}{\cal E}_{1-}) -  s(\lambda_{1}\Gamma^{a} M \omega_{2})
 = (\lambda_{1}\Gamma^a)_{\beta} [{\cal E}_{1-}^{\beta} +
(M d_{2})^{\beta} - \tilde C_{2 \alpha}{}^{\beta \gamma} 
\lambda_{2}^{\alpha}\omega_{2 \gamma}] = 0,
\label{65a}
\end{eqnarray}
\begin{eqnarray}
(\lambda_{2}\Gamma^{ a}{\cal E}_{2 +})+ s(\lambda_{2}\Gamma^{a} M \omega_{1}) = 
(\lambda_{2}\Gamma^a)_{\beta}[{\cal E}_{2 +}^{\beta} - ( d_{1} M )^{ \beta} + 
\tilde C_{1 \alpha}{}^{
 \beta \gamma} \lambda_{1}^{\alpha}\omega_{1 \gamma}] = 0,
\label{65b}
\end{eqnarray}
where 
\begin{eqnarray}
\tilde C_{i \alpha}{}^{\beta \gamma} =  C_{i \alpha}{}^{\beta \gamma} +
Y_{i \alpha}{}^{\beta \gamma},
\label{65c}
\end{eqnarray}
and $ C_{i \alpha}{}^{\beta \gamma}$ and $ Y_{i \alpha}{}^{\beta \gamma}$
are defined in (\ref{45b}) and (\ref{53c}).
   
The failure of nilpotency in equations (\ref{63a}) and (\ref{63b}) is a 
consequence of the 
$\omega$-gauge transformation (\ref{47}). Indeed $s^2$,
acting on $\omega$, vanishes only modulo this gauge transformation.

One can cure this inconvenience by fixing the $\omega$-gauge and a useful
way to do that is to apply the so-called Y-formalism. 

Given the constant spinors $V_{\underline{\alpha}} = (V_{1 \alpha},
 V_{2 \alpha})$ one defines
\begin{eqnarray}
K_{\underline{\alpha}}{}^{\underline{\beta}} = (K ^{(1)}_{\alpha}{}^{\beta},
\quad K ^{(2)}_{ \alpha}{}^{ \beta}),
\nonumber
\end{eqnarray}
where 
\begin{eqnarray}
K ^{(1)}_{\alpha}{}^{\beta} = {1 \over 2} (\Gamma^{a}\lambda_{1})_{\alpha}
(Y_{1}\Gamma_{a})^{\beta},
\nonumber
\end{eqnarray}
\begin{eqnarray}
K^{(2)}_{\alpha}{}^{\beta} = {1 \over 2} (\Gamma^{a}\lambda_{2})_{\alpha}
(Y_{2}\Gamma_{a})^{ \beta},
\label{66}
\end{eqnarray}
%%%%%%%%%%%%%%%%%%%%%
and $$ Y_{i} = { {V_{i}} \over
{(V_{i} \lambda_{i}})}, $$    so that $$  (Y_{i}\lambda_{i}) = 1 .$$
Moreover
%%%%%%%%%%%%%%%%%
\begin{eqnarray}
( \lambda_{1}K^{(1)})^{\alpha} = 0 = (\lambda_{2} K^{(2)} )^{\alpha},
\label{67}
\end{eqnarray}
%%%%%%%%%%%%%%%%%%%%%
and
%%% %%%%%%%%%%%%%%%
\begin{eqnarray}
((1 - K^{(1)})\Gamma^{a}\lambda_{1})_{\alpha} = 0 =
((1 - K^{(2)})\Gamma^{a}\lambda_{2})_{\alpha}.
\label{68}
\end{eqnarray}
Here $K^{(1)}$ and $K^{(2)}$ are projectors and, since $Tr K_{i} = 5 $, they 
project on  five-dimensional subspaces of the 16-dimensional spinorial spaces 
so that from (\ref{59}) one can see that $\lambda_{1}$ and $\lambda_{2}$ have
11 independent components.

Using the projectors $K^{(i)}$ one can fix the $\omega$-gauge symmetry by 
requiring 
\begin{eqnarray}
(K^{(1)}\omega_{1})_{\alpha} = 0 = (K^{(2)}\omega_{2})_{\alpha},
\label{69}
\end{eqnarray}
or equivalently, 
\begin{eqnarray}
\omega_{i} = ((1 - K^{(i)})\omega_{i}),
\label{70}
\end{eqnarray}
so that each of the $\omega_{i}$  also has 11 components.
Moreover one can also split the fields $d_{\underline \alpha}$ as
$$ d_{i}^{(\top)} = ((1 - K^{(i)})d_{i}, $$ 
$$ d_{i}^{(\bot)} = K^{(i)} d_{i}.$$ 
Notice that only $ d_{i}^{(\top)} $ appear in the BRST charge $Q$ so that $ 
d_{i}^{(\top)}$ are the BRST partners of
$\omega_{i}$.    
Since $ V_{i} $ are constants, $ K^{(i)} $ break  Lorentz 
invariance, and are singular at $ (V_{i} \lambda_{i}) = 0 $ but these facts are not 
a problem since, as we will see, any dependence on  $ K^{(i)} $ disappears in
the final result.
 
Projecting (\ref{57}), (\ref{59}), with $ (1 - K)$ one gets
 $s \omega_{i}$ and $ s d^{(\top)}_{i}$ which,
 in covariant form are
\begin{eqnarray}
 \tilde s \omega_{\underline{\alpha}} = - d^{(\top)}_{\underline{\alpha}} -
\lambda^{\underline{\beta}} X_{\underline{\beta}\underline{\alpha}}{}^{
\underline{\gamma}}\omega_{\underline{\gamma}},
\label{107}
\end{eqnarray}
\begin{eqnarray}
 \tilde s  d^{(\top)}_{\underline{\alpha}} = \lambda^{\underline{\beta}} X_{
\underline{\beta}\underline{\alpha}}{}^{\underline{\gamma}}d^{(\top)}_{
\underline{\gamma}} +\lambda^{\underline{\delta}}  \lambda^{\underline{\beta}}
 \tilde R_{
\underline{\alpha}\underline{\delta}\underline{\beta}}{}^{\underline{\gamma}}
\omega_{\underline{\gamma}},
\label{108}
\end{eqnarray}
and projecting (\ref{63a}), (\ref{63b}), (\ref{64}) and 
(\ref{65}) with $ (1 - K)$ one has
\begin{eqnarray}
 s^{2} \omega_{\underline{\alpha}} = 0 = s^{2} d^{(\top)}_{\underline{\alpha}}.
\label{109}
\end{eqnarray}
As for $ s d^{\bot}$,  notice that, given the definition (\ref{54}) of 
$d_{\underline{\alpha}}$, only the components of $  \tilde \Omega_{
\underline{\alpha} \underline{\beta}}{}^{\underline{\gamma}}$ projected
with $ (1 - K)_{\underline{\beta'}}^{\underline{\beta}}$ and $
 (1 - K)_{\underline{\gamma}}^{\underline{\gamma'}}$ are present in (\ref{59})
 so that
\begin{eqnarray}
\tilde s d^{(\bot )}_{1 \alpha} = - (\Gamma_{a}\lambda_{1})_{\alpha}
 [ {\cal E}_{+}^{a} + (\omega_{1} M \Gamma^{a} \lambda_{2})], 
\label{110}
\end{eqnarray}
\begin{eqnarray}
\tilde s d^{(\bot )}_{2 \alpha} = - (\Gamma_{a}\lambda_{2})_{\alpha}
 [ {\cal E}_{-}^{a} - (\lambda_{1} \Gamma^{a}M \omega_{2})]. 
\label{111}
\end{eqnarray}
 Moreover $ s^2 d^{\bot }_{\underline{\alpha}}$ is just given by
(\ref{64a}) and (\ref{64b}) which,  as we will see,  vanishes on shell.
 Now $ s $ is  nilpotent acting on any field or ghost.

\subsection{\bf  Derivation of the Action}

In {\cite{Ton}} two methods are presented to derive the pure spinor 
action in IIA superstring $\sigma$-models. Both methods can be applied to 
the present case. In this subsection we will give the details of only the 
second method first proposed in \cite{Oda1} for  heterotic $\sigma$-models.

The strategy is the following: as a first step one adds to the Green-Schwarz  
action $I_{GS}$ a new action $I_{K}$ that depends on the projector 
$ K_{\underline{\alpha}}{}^{\underline{\beta}}$ such that the action 
$I_{GS} + I_{K}$ is BRST invariant; then one adds a BRST exact action term
$I_{gf}$ (the "gauge fixing'' action),  given by the BRST transformation of a 
``gauge fermion'' with ghost number $n_{gh} = - 1 $ such that the 
BRST invariant action $I = I_{GS} + I_{K} + I_{gf}$ becomes 
independent of  $ K_{\underline{\alpha}}{}^{\underline{\beta}}$.

The Green-Schwarz action is given in (\ref{105}) and its BRST transformation 
in (\ref{106}). Now consider the action $$ I_{K} = I_{K}^{(1)} +  
I_{K}^{(2)} + I_{K}^{(3)},$$ 
where 
\begin{eqnarray}
 I_{K}^{(1)} = \oint [ ( {\cal E}_{1 -}
K^{(1)}d_{1}) + ({\cal E}_{2 +} K^{(2)}d_{2})], 
\label{112a}
\end{eqnarray}
\begin{eqnarray}
 I_{K}^{(2)} = - \oint (d_{1}\tilde K^{(1)} M K^{(2)}d_{2}), 
\label{112b}
\end{eqnarray}
\begin{eqnarray}
 I_{K}^{(3)} = \oint (\omega_{1} M\Gamma_{a}\lambda_{2})(\lambda_{1}\Gamma^{
a} M \omega_{2}).
\label{112c}
\end{eqnarray}
with $ \tilde K^{(i)}$ being the transpose of $K^{(i)}$. 
Notice that equations (\ref{102}) and (\ref{103}), projected with $K$ reduce to 
\begin{eqnarray}
 \tilde s ({\cal E}_{1} K^{(1)}))^{\alpha} =   - {\cal E}^{c}(M \Gamma_{c}
\lambda_{2})^{\alpha}, 
\label {112}
\end{eqnarray}
\begin{eqnarray}
  \tilde s ({\cal E}_{2}K^{(2)})^{\alpha} =   
 {\cal E}^{c}(\lambda_{1} \Gamma_{c} M )^{\alpha}. 
\label{113}
\end{eqnarray}
An explicit computation of $ s I_{K}$, using equations (\ref{110}), (\ref{111}),
(\ref{112}) and (\ref{113}) as well as (\ref{43c}) and (\ref{107}) and taking 
into account equations (\ref{45a}) and (\ref{45b}) yields
\begin{eqnarray}
s I_{K} = - \oint  [(\lambda_{1}{\cal E}_{+}^{a}\Gamma_{a}{\cal E}_{1 -}) + 
(\lambda_{2}{\cal E}_{-}^{a}\Gamma_{a}{\cal E}_{2 +})] 
\nonumber
\end{eqnarray}
\begin{eqnarray}
 - \oint (\omega_{1} M \Gamma^{a} \lambda_{2})(\lambda_{1}
\Gamma_{a})_{\beta}  [{\cal E}_{1 -}^{\beta} + 
(M d_{2})^{\beta} - C_{2  \alpha}{}^{\beta  \gamma} 
\lambda_{2}^{ \alpha}\omega_{2 \gamma}]
\nonumber
\end{eqnarray}
\begin{eqnarray}
+ \oint [{\cal E}_{2}^{
 \beta} - ( d_{1} M )^{ \beta} + C_{1 \alpha}{}^{
 \beta \gamma} \lambda_{1}^{\alpha}\omega_{1 \gamma}](\Gamma^{a}\lambda_{2})_{
\beta}(\lambda_{1}\Gamma_{a}M\omega_{2}), 
\label{114}
\end{eqnarray}
where the last two integrals vanish on shell, as discussed before (see 
(\ref{65a}), (\ref{65b})) and proved later on.
Therefore, given (\ref{106}),$$ s I_{GS} + I_{K} = 0, $$
on shell.

Now we define $I_{gf}$ as
\begin{eqnarray}
I_{gf} = - s  \oint [({\cal E}_{1-} \omega_{1}) + ({\cal E}_{2 +} \omega_{2})]
  + s \oint [(d_{1}\tilde K^{(1)} M \omega_{2}) - (\omega_{1} M K^{(2)}d_{2})]
\nonumber
\end{eqnarray}
\begin{eqnarray}
- {1 \over 2} s \oint (s_{1} -  s_{2}) (\omega_{1} M \omega_{2}). 
\label{115}
\end{eqnarray}
Performing the BRST variations one has
\begin{eqnarray}
- s  \oint [({\cal E}_{1 -} \omega_{1}) + {\cal E}_{2 +} \omega_{2})] =
(\omega_{1} \tilde \Delta_{-}\lambda_{1}) + (\omega_{2} \tilde \Delta_{+}
\lambda_{2}) +  {\cal E}_{-}^{a}(\omega_{1} M \Gamma_{a}\lambda_{2}) -
{\cal E}_{+}^{a}( \lambda_{1}\Gamma_{a}M\omega_{2}) 
\nonumber
\end{eqnarray}
\begin{eqnarray}
+ {\cal E}_{1 -} ((1 - K^{(1)})d_{1}) +  {\cal E}_{2 +} ((1 - K^{(2)})d_{2}),  
\label{116} 
\end{eqnarray}
\begin{eqnarray}
 + s \oint [(d_{1}\tilde K^{(1)} M \omega_{2}) - (\omega_{1} M K^{(2)}d_{2})]=
{\cal E}_{+}^{a}( (\lambda_{1}\Gamma_{a}M\omega_{2}) -  {\cal E}_{-}^{a}(
\omega_{1} M \Gamma_{a}\lambda_{2}) 
\nonumber
\end{eqnarray}
\begin{eqnarray}
+ (d_{1}K^{(1)})_{\alpha}\tilde C^{
\alpha \gamma}_{2 \beta}\lambda_{2}^{\beta}\omega_{2 \gamma} 
+ \omega_{1 \gamma}\lambda_{1 \beta}\tilde C_{1 \beta}^{\gamma \alpha} (
K^{(2)}d_2)_\alpha - ( d_{1}K^{(1)}M(1 - K^{(2)}) d_{2}) 
\nonumber
\end{eqnarray}
\begin{eqnarray}
-( d_{1}(1 - K^{(1)})M K^{(2)} d_{2})
- 2 (\omega_{1} M\Gamma_{a}\lambda_{2})(\lambda_{1}\Gamma^{
a} M \omega_{2}),
\label{117}
\end{eqnarray}
\begin{eqnarray}
- {1 \over 2} s \oint (s_{1} -  s_{2}) (\omega_{1} M \omega_{2})= 
 (d_{1}(1 - K^{(1)}))_{\alpha}\tilde C^{\alpha \gamma}_{2 \beta}
\lambda_{2}^{\beta}\omega_{2 \gamma} + \omega_{1 \gamma}\lambda_{1 \beta}
\tilde C_{1 \beta}^{\gamma \alpha}((1 - K^{(2)})d_2)_\alpha 
\nonumber
\end{eqnarray}
\begin{eqnarray}
- ( d_{1}(1 - K^{(1)})M(1 - K^{(2)}) d_{2}) +
\omega_{1 \beta}\lambda_{1}^{\alpha}S_{\alpha \gamma}^{\beta \delta}
\lambda_{2 \gamma}\omega_{2 \delta} + (\omega_{1} M\Gamma_{a}\lambda_{2})
(\lambda_{1}\Gamma^{a} M \omega_{2}),
\label{118}
\end{eqnarray}
where
\begin{eqnarray}
S_{\alpha \gamma}^{\beta \delta} ={1 \over 2} C_{\alpha \gamma}^{\beta \delta}
 - (Y^{(1)\beta}_{\alpha \eta}\tilde C^{\eta \delta}_{2  \gamma} - 
C_{1 \alpha}^{\beta \eta}Y^{(2) \delta}_{\eta \gamma}) + Y^{(1) \beta}_{\alpha\eta}
M^{\eta \kappa} Y^{(2) \delta}_{\kappa \gamma},
\label{119}
\end{eqnarray}
and 
\begin{eqnarray}
\lambda_{1}^{\alpha}\lambda_{2}^{\gamma}C_{\alpha \gamma}^{\beta \delta} =
s_{2}s_{1}P^{\beta \delta} = - s_{1}s_{2}P^{\beta \delta}.
\label{120}
\end{eqnarray}
It follows from (\ref{45b}) that the fields  $C_{\alpha \gamma}^{\beta \delta}
$, and therefore $ S_{\alpha \gamma}^{\beta \delta}$, are Lorentz-Weyl valued 
in $\alpha$, $\beta$ and in $\gamma$, $\delta$.

Adding equations (\ref{105}), (\ref{112a}), (\ref{112b}), (\ref{112c}) and
(\ref{115}), one obtains the pure spinor action:
\begin{eqnarray}
I = I_{GS} + I_{K} + I_{gf} =  \int [{1 \over 2}{\cal E}_{+}^{a}{\cal E}_{
- a } + ( n_{i}B_{2}^{i})+ (\omega_{1} \tilde \Delta_{+}\lambda_{1}) + 
(\omega_{2} \tilde \Delta_{-}\lambda_{2})  +  ({\cal E}_{1 +}d_{1}) + 
({\cal E}_{2 -}d_{2})  
\nonumber
\end{eqnarray}
\begin{eqnarray}
- ( d_{1} M d_{2}) +  d_{1  \alpha}\tilde C^{\alpha \gamma}_{2 \beta}
\lambda_{2}^{\beta}\omega_{2 \gamma} + \omega_{1 \gamma}\lambda_{1 \beta}
\tilde C_{1 \beta}^{\gamma \alpha}d_{2 \alpha}  +
\omega_{1 \beta}\lambda_{1}^{\alpha}S_{\alpha \gamma}^{\beta \delta}
\lambda_{2 \gamma}\omega_{2 \delta}],
\label{121}
\end{eqnarray}
which is in full agreement with the pure spinor action first obtained by Berkovits and
 Howe in \cite{BH}.

Notice that the field equations obtained from this action varying 
$d_{\underline{\alpha}}$ are 
$$ {\cal E}_{1}^{\beta} +
(M d_{2})^{\beta} - \tilde C_{2 \alpha}^{\beta \gamma} 
\lambda_{2}^{\alpha}\omega_{2 \gamma} = 0, $$ $$ {\cal E}_{2}^{\beta} -
( d_{1} M )^{ \beta} + \tilde C_{1 \alpha}^{
 \beta \gamma} \lambda_{1}^{\alpha}\omega_{1 \gamma} = 0, $$
which justify equations (\ref{65a}) and (\ref{65b}) and assure the on shell 
nilpotence of $s$ acting on $d_{\underline{\alpha}}$.

%%%%%%%%%%%%%%%%%%%%%%%%%%%%%%%%%%%%%%%%%%%%%%%%%%%%%%%%%%%%%%%%%%
%%%%%%%%%%%%%%%%%%%%%%%% Acknowledgements %%%%%%%%%%%%%%%%%%%%%%%%%%%%%
%%%%%%%%%%%%%%%%%%%%%%%%%%%%%%%%%%%%%%%%%%%%%%%%%%%%%%%%%%%%%%%%%%
\begin{flushleft}
{\bf Acknowledgements}
\end{flushleft}

This work (I.O.) is supported in part by the Grant-in-Aid for Scientific 
Research (C) No. 22540287 from the Japan Ministry of Education, Culture, 
Sports, Science and Technology.

\end{document}